\documentclass[fleqn,usenatbib]{mnras}
\makeatletter
\def\caption{\H@refstepcounter\@captype\@dblarg{\@caption\@captype}}
\makeatother

\usepackage{newtxtext,newtxmath}
\usepackage{xspace}
\usepackage{bm}
\usepackage{xcolor}

\usepackage[T1]{fontenc}
\usepackage{booktabs}
\usepackage{cleveref}
\usepackage{tabularx}
\usepackage{siunitx}
\usepackage{textcomp}
\usepackage{orcidlink}
\crefname{figure}{Fig.}{Figs}
\Crefname{figure}{Fig.}{Figs}
\crefname{table}{Table}{Tables}
\crefname{equation}{equation}{equations}
\Crefname{equation}{Equation}{Equations}
\creflabelformat{equation}{(#2#1#3)}
\crefname{section}{Section}{Sections}
\crefname{appendix}{Appendix}{Appendices}
\usepackage{enumitem}

\DeclareRobustCommand{\VAN}[3]{#2}
\let\VANthebibliography\thebibliography
\def\thebibliography{\DeclareRobustCommand{\VAN}[3]{##3}\VANthebibliography}

\usepackage{graphicx}	
\usepackage{amsmath}	

\newcommand{\arepo}{{\scshape arepo}\xspace}
\newcommand{\nori}{{\scshape nori}\xspace}
\newcommand{\gizmo}{{\scshape gizmo}\xspace}
\newcommand{\arepoOne}{{\scshape arepo-1}\xspace}
\newcommand{\arepoTwo}{{\scshape arepo-2}\xspace}

\newcommand\orcid[1]{\orcidlink{#1}}
\newcommand{\appropto}{\mathrel{\vcenter{
  \offinterlineskip\halign{\hfil$##$\cr
    \propto\cr\noalign{\kern2pt}\sim\cr\noalign{\kern-2pt}}}}}

\title[SIDM on GPUs]{\nori: a GPU-accelerated code for simulating self-gravitating and self-interacting dark matter}

\author[Zier et al.]{Oliver Zier\orcid{0000-0003-1811-8915},$^{1}$\thanks{E-mail: oliver.zier@cfa.harvard.edu}
R\"udiger Pakmor\orcid{0000-0003-3308-2420},$^{2}$
Xuejian Shen\orcid{0000-0002-6196-823X},$^{1,3,4}$
Rongrong Liu\orcid{0000-0003-0685-3525},$^{1}$
Lucas Schleuss\orcid{0009-0000-0276-6622},$^{2}$
\newauthor
 Mark Vogelsberger\orcid{0000-0001-8593-7692},$^{3,4}$ and Lars Hernquist\orcid{0000-0001-6950-1629},$^{1}$
\\
$^{1}$ Center for Astrophysics $|$ Harvard $\&$ Smithsonian, 60 Garden Street, Cambridge, MA 02138, USA\\
$^{2}$ Max-Planck-Institut f\"ur Astrophysik, Karl-Schwarzschild-Str.~1, D-85748 Garching, Germany\\%
$^{3}$ Massachusetts Institute of Technology, Kavli Institute for Astrophysics and Space Research, 77 Massachusetts Avenue, Cambridge, MA 02139, USA\\%
$^{4}$ Department of Physics, Massachusetts Institute of Technology, 77 Massachusetts Avenue, Cambridge, MA 02139, USA\\
}

\date{Accepted XXX. Received YYY; in original form ZZZ}

\pubyear{\the\year{}}

\begin{document}
\label{firstpage}
\pagerange{\pageref{firstpage}--\pageref{lastpage}}
\maketitle

\begin{abstract}
Self-interacting dark matter (SIDM) can form cores in dark matter haloes and later drive their gravothermal collapse, raising their central densities far above those of NFW haloes. Simulating this collapse is computationally demanding, requiring increasingly short time-steps and a large particle number to resolve the shrinking core. We present \nori, an $N$-body code that runs on CPUs and GPUs and simulates isolated self-gravitating systems with elastic, isotropic self-interactions using a Monte-Carlo method. To avoid MPI communication in the tree walks, each compute node in CPU mode, and each GPU or group of GPUs in GPU mode, holds a copy of all particles and its own Morton-ordered tree. A fixed evaluation order makes the results bitwise identical on CPUs and GPUs. Applying scattering events sequentially conserves the momentum and energy of each pair to round-off and allows multiple scatterings per particle per step. With global time-steps in both codes, \nori is several times faster on CPUs than our \arepoTwo implementation. One four-A100 GPU node is up to 60 times faster per time-step than \arepoTwo on a 112-core CPU node. For a full core-collapse simulation, where \arepoTwo uses individual time-steps, the speed-up is $\sim 30$ for the central density to reach $10^3$--$10^4$ times the NFW scale density, $\rho_{\rm s}$. We follow a halo with $10^8$ particles within $r_{200}$ to a central density of $10^4\,\rho_{\rm s}$. Within the limit of about $2\times10^9$ particles set by data replication, \nori provides an efficient tool for isolated SIDM simulations, including gravothermal collapse and cluster collisions.
\end{abstract}



\section{Introduction}

In the standard $\Lambda$CDM cosmological model, cold dark matter (CDM) is non-relativistic, does not
interact with electromagnetic radiation, and is effectively
collisionless. The CDM model provides a successful description of
structure formation on cosmological scales. On galactic and sub-galactic
scales, however, comparisons between gravity-only CDM simulations and
observations have revealed several possible tensions. These include the
``cusp--core'' and ``missing satellites'' problems, the finding that the most massive predicted subhaloes are too dense to host the observed bright satellites (``too big to fail''), and the
diversity of dwarf-galaxy rotation curves
\citep{Flores1994,Moore1994CuspCore,Klypin1999MissingSatellites,
BoylanKolchin2011TBTF,BoylanKolchin2012,Oman2015Diversity}. The significance of these tensions
remains actively debated, since baryonic processes, selection effects, and
uncertainties in inferring halo properties can alleviate at least some of
them \citep{BullockBoylanKolchin2017SmallScaleReview}.

One widely studied solution to these tensions is self-interacting dark matter (SIDM), in
which dark matter particles undergo non-gravitational scattering
\citep{SpergelSteinhardt2000SIDM}. Repeated scattering transports energy
through a halo, providing an effective mechanism for heat
conduction. During the early evolution, heat flows from the dynamically
hotter outer halo into the colder central cusp, lowering the central density and producing an
approximately isothermal core. At the relatively low velocities of dwarf galaxies, cross-sections of $\sigma/m\sim1\text{--}10\,\mathrm{cm^2\,g^{-1}}$ produce kpc-sized cores and substantially lower the central densities of haloes, potentially alleviating some of the tensions discussed above \citep{vogelsberger2012subhaloes,rocha2013cosmological,Elbert2015SIDMCores,KaplinghatTulinYu2016ParticleColliders,TulinYu2018SIDMReview}.
In galaxy clusters, however, where velocities are much higher, the density profiles and shapes of cluster haloes and merging clusters constrain the cross-section to considerably smaller values \citep{Randall2008BulletCluster,Peter2013SIDMHaloShapes,KaplinghatTulinYu2016ParticleColliders,Sagunski2021}.
This motivates velocity-dependent models in which the interaction is strong at low velocities but suppressed at high velocities.
A velocity-dependent cross-section is also natural for interactions mediated by a light
force carrier \citep{Feng2010,BuckleyFox2010LightMediators,LoebWeiner2011YukawaSIDM,TulinYuZurek2013SIDM}.

Core formation is not the final stage of this evolution. Once the
isothermal core has formed, the central region can become
hotter than the surrounding halo, which reverses the heat flux. Because a
self-gravitating system has a negative heat capacity, the loss of energy
causes the core to contract and heat up further, resulting in a runaway gravothermal collapse
\citep{Lynden1968,Lynden-Bell1980,Balberg2002,KodaShapiro2011Gravothermal}.
At high
redshift, gravothermal collapse has been proposed as a route to forming seeds of
supermassive black holes in elastic and dissipative SIDM models \citep{Balberg2002,Pollack2015,Feng2021,Xiao2021,Shen2025,Jiang2026,Roberts2026}.
On subhalo scales, core-collapsed objects may instead account for unusually
dense perturbers inferred from strong gravitational lensing arcs~\citep{Vegetti2010,Meneghetti2020,Minor2021,Vegetti2026,Powell2025,Enzi2025}, galaxy-galaxy strong lensing cross-sections~\citep{Dutra2025,Natarajan2026}, and perturbers of cold stellar
streams \citep{Zhang2024GD1SIDM,Yu2025}. Both applications
require reliable predictions for the late stages of collapse, during which the
central density rises by several orders of magnitude.

There are several approaches to modelling this evolution. Fluid models approximate the halo as a conducting fluid.
They are inexpensive enough to follow the collapse to very high densities,
but need a closure relation for the heat flux or a calibrated conductivity
\citep{Balberg2002,KodaShapiro2011Gravothermal,Correa2021,Outmezguine2023,Yang2023}. Three-dimensional $N$-body simulations instead represent the collision term by random scattering events between neighbouring simulation particles \citep{Burkert2000,Yoshida2000,Dave2001SIDM,vogelsberger2012subhaloes, rocha2013cosmological,Robertson2017SIDM,Correa2022}. Alternatively, a drag--diffusion method can be used for frequent interactions that are strongly peaked in the forward direction \citep{Fischer2021,Fischer2024}. Intermediate approaches include parametric halo models calibrated on $N$-body simulations \citep{Yang2024} and spherically symmetric particle codes with stochastic scattering \citep{Kamionkowski2025}, while \citet{Ramos2025} derived an effective interaction between simulation particles from the microphysics in the short-mean-free-path regime. Three-dimensional particle simulations retain
the full phase-space information and can treat systems without assuming
spherical symmetry or isolation. They are also used to calibrate the simpler models
\citep{Outmezguine2023,Yang2024,Mace2026}. Their numerical accuracy therefore
directly affects the predictions made with these models.

Accurate particle simulations become particularly difficult once the halo
starts to collapse and the scattering frequency in the core grows.
Individual scattering events have to be resolved in time, which imposes strict limits on the time-steps of Monte Carlo schemes.
The computational cost can be reduced by integrating particles on individual time-steps, though this can introduce energy errors \citep{Fischer2024Energy}.
The collapsing core also loses mass and contains fewer and fewer particles.
 A large initial particle number is therefore required to resolve the collapse to high densities.

Prominent examples of general-purpose codes with SIDM implementations are
\textsc{gadget-2} \citep{Springel2005Gadget2,rocha2013cosmological},
\textsc{OpenGadget3} \citep{OpenGadget3,Fischer2026OpenGadget3SIDM},
\gizmo \citep{Hopkins2015GIZMO,robles2017sidm,Shen2021dSIDM,meskhidze2022comparing},
\textsc{swift} \citep{Correa2022,Schaller2024SWIFT}, and
\arepo \citep{springel2010pur,Vogelsberger2014,vogelsberger2019evaporating}, whose current version, \arepoTwo, contains our recent implementation \citep{ZierSidm2026}; we refer to the earlier implementation \citep{Vogelsberger2014,vogelsberger2019evaporating} as \arepoOne.
These codes use spatial domain decomposition and individual particle time-steps, which are well suited to large cosmological simulations and offer considerable flexibility.

Graphics processing units (GPUs) have accelerated collisionless $N$-body codes for more than a decade.
\textsc{Bonsai} runs all parts of the tree code, including the tree construction, on the GPU \citep{Bedorf2012} and has been scaled to 18\,600 GPUs \citep{Bedorf2014Bonsai}. \textsc{pkdgrav3} evaluates the particle--particle and particle--cell interactions of its fast multipole method on GPUs and has run a two-trillion-particle simulation \citep{Potter2017PKDGRAV3}. \textsc{hacc} computes its short-range force on GPUs \citep{Habib2016HACC}; and \textsc{abacus} computes the near-field force by direct summation on GPUs \citep{Garrison2021Abacus}.
All four distribute the particles over the compute nodes by a spatial domain decomposition, so that data from neighbouring domains must be exchanged or duplicated. Once the GPUs do the arithmetic, this communication can become the main barrier to scaling \citep{Bedorf2014Bonsai}.
\textsc{OpenGadget3} offloads gravity, hydrodynamics, and thermal conduction to GPUs \citep{Ragagnin2020GPU,Ragagnin2026OpenGadget3GPU}, but its SIDM module cannot currently be combined with the offloading \citep[their table~7]{OpenGadget3}, and SIDM simulations have so far relied on CPUs only \citep{Fischer2026OpenGadget3SIDM}.

Isolated SIDM simulations typically contain fewer than $10^8$ particles, so the whole system fits into the memory of one compute node, or even of one GPU.
This permits a simpler parallelization without spatial domain decomposition, in which every node or GPU holds a complete copy of the system.
We adopt this replicated-data approach in \nori\footnote{\nori stands for $N$-body code Optimized for
Random Interactions.}, a new $N$-body code for isolated systems of equal-mass
self-interacting dark matter particles and, to our knowledge, the first SIDM code that runs on GPUs.
It supports a shared-memory CPU mode as well as a GPU mode,
both using a one-sided gravity tree walk on a tree built from the bottom up.
The tree is replicated as well, so that the tree walks need no MPI communication and the work can be divided freely between CPU cores or GPUs.
The replication limits the maximum particle number to around $2\times10^{9}$.

The paper is organized as follows. In \cref{sec:methods}, we first summarize
the numerical methods required for an SIDM $N$-body simulation and discuss
their main sources of error. We describe the implementation of \nori in
\cref{sec:nori}. In \cref{sec:validation}, we test the force accuracy and the
long-term stability of a collisionless halo before comparing an SIDM core
collapse with our \arepoTwo implementation. We then study the CPU and GPU performance in
\cref{sec:performance} and introduce a higher-resolution reference
calculation in \cref{sec:referenceSolution}.  Finally, we discuss the
limitations of the code in \cref{sec:limitations} and summarize our results in
\cref{sec:conclusions}.

\section{Methods for simulating self-interacting and self-gravitating dark matter}
\label{sec:methods}
This section summarizes the numerical ingredients of an isolated SIDM $N$-body simulation and the errors associated with each of them.

\subsection{Particle representation}
The evolution of self-interacting dark matter particles in phase space is described by
the collisional Boltzmann equation
\begin{equation}
    \frac{\partial f}{\partial t}
    +\bm{v}\cdot\nabla_{\bm{x}}f
    -\nabla_{\bm{x}}\Phi\cdot\nabla_{\bm{v}}f
    =
    \mathcal{C}\!\left[f;\frac{\mathrm{d}\sigma}{\mathrm{d}\Omega}\right],
    \label{eq:CBE}
\end{equation}
where $f(\bm{x},\bm{v},t)$ is the phase-space mass density
and $\Phi$ denotes the gravitational potential. The collision operator
$\mathcal{C}$ depends on the differential cross-section
$\mathrm{d}\sigma/\mathrm{d}\Omega$ of the microscopic particles. Integrating
$f$ over velocity space yields the mass density
$\rho=\int f\,\mathrm{d}^3\bm{v}$, which determines the gravitational potential through Poisson's equation:
\begin{equation}
    \nabla^2\Phi=4\pi G\rho.
      \label{eq:Poisson}
\end{equation}
Here, $G$ is the gravitational constant. Without
self-interactions, $\mathcal{C}=0$, \cref{eq:CBE} becomes the
collisionless Boltzmann (Vlasov) equation.

An $N$-body scheme represents $f$ by $N$ simulation particles with masses $m_i$, all equal to $m_{\rm p}$ in this paper,
positions $\bm{x}_i$, and velocities $\bm{v}_i$:
\begin{equation}
    f_N(\bm{x},\bm{v},t)
    = \sum_{i=1}^{N}m_i\,
      \delta^{(3)}(\bm{x}-\bm{x}_i)\,
      \delta^{(3)}(\bm{v}-\bm{v}_i).
    \label{eq:particle_df}
\end{equation}
The simulation particles sample the distribution function and do not correspond
to individual microscopic dark matter particles. Between scattering events, their
motion is governed by
\begin{equation}
    \frac{\mathrm{d}\bm{x}_i}{\mathrm{d}t} = \bm{v}_i,
    \qquad
    \frac{\mathrm{d}\bm{v}_i}{\mathrm{d}t}
    = -\nabla\Phi_\epsilon(\bm{x}_i),
    \label{eq:eom}
\end{equation}
where we use the softened collective potential
\begin{equation}
    \Phi_\epsilon(\bm{x})
    = -G\sum_j m_j\,\varphi_\epsilon(|\bm{x}-\bm{x}_j|),
    \label{eq:softened_potential}
\end{equation}
where $\varphi_\epsilon$ is the softened pair potential given in \cref{app:SofteningKernel}.
Equivalently, $\Phi_\epsilon$ is the potential of a smoothed density in which the mass of each particle is spread over the cubic-spline kernel introduced in \cref{subsec:sidm_methods} with radius $h_\epsilon=2.8\epsilon$.

Gravitational softening reduces force fluctuations caused by finite particle
sampling and suppresses strong deflections during close encounters between
simulation particles \citep{Dehnen2001,Power2003}. 
Both effects are discreteness
artefacts: in the collisionless limit of \cref{eq:CBE}, each particle should
respond only to the smooth potential of the full distribution. Softening
removes the strongest encounters but not the force noise of finite sampling.
The softening length sets
a numerical force-resolution scale and does not represent the physical
self-interactions described by $\mathcal{C}$. Its choice balances
discreteness noise against systematic force errors: too little softening
leaves large force fluctuations, and excessive softening biases the
gravitational force in the halo centre. In this work, we use the
cubic-spline gravitational softening of the \textsc{gadget} family
\citep{springel2001gadget,Springel2005Gadget2}. The force becomes exactly
Newtonian beyond $2.8\epsilon$, where $\epsilon$ denotes the
Plummer-equivalent softening length quoted throughout this paper.
In this paper, all particles share the same softening length. It would also be possible to extend this scheme to individual particle softening lengths, though this requires the pair potential to be symmetric in the two softening lengths to conserve energy and momentum. A possible choice is $\max(\epsilon_i,\epsilon_j)$.

Even with softened gravity, finite particle sampling introduces statistical
fluctuations and artificial gravitational relaxation. Increasing the
particle number at fixed halo mass reduces these effects, but the
resolution requirements become increasingly demanding during gravothermal
core collapse. As the core contracts and loses mass, fewer simulation
particles sample its structure, while its shrinking radius becomes less
well resolved by a fixed softening length. A simulation that adequately
resolves the initial halo can therefore become under-resolved at later
times.

\subsection{Approximate self-gravity}
\label{sec:gravity_methods}
For vacuum (non-periodic) boundary conditions, a direct summation of the gradient of \cref{eq:softened_potential} provides the natural reference solution for the gravitational force. Its $\mathcal{O}(N^2)$ cost makes it impractical for repeated force evaluations in
large simulations, although it remains useful as ground truth for accuracy tests.
Mesh-based methods reduce this cost by assigning the particle masses to a
grid and solving the discretized Poisson equation, either using Fourier
transforms \citep{Hockney1988nbody} or real-space methods such as multigrid
\citep{guillet2011simple}. On a uniform grid, the force resolution is limited by
the grid spacing; efficiently resolving a wide range of spatial scales
therefore requires mesh refinement or a hybrid method such as TreePM
\citep{Bagla2002}.
Tree methods instead adapt to
the particle distribution \citep{Barnes1986}. In an octree, nearby particles
are combined into cells.
Sufficiently distant cells can then be represented
by a multipole expansion around their centre of mass.
 The monopole is the total cell mass, the dipole vanishes about the centre of mass, and the first correction is the quadrupole.
Higher moments can also be used \citep[e.g.][]{springel2021simulating}.
Increasing the multipole order improves the accuracy but also increases the memory consumption and cost per interaction.
The tree can be constructed by inserting particles one by one and splitting cells that contain more than one particle into eight subcells.
We call a subdivided cell an internal node and a cell that is not subdivided further a leaf.
Allowing more than one particle per leaf reduces the depth of the tree.

\begin{figure*}
    \centering
    \includegraphics[width=\textwidth]{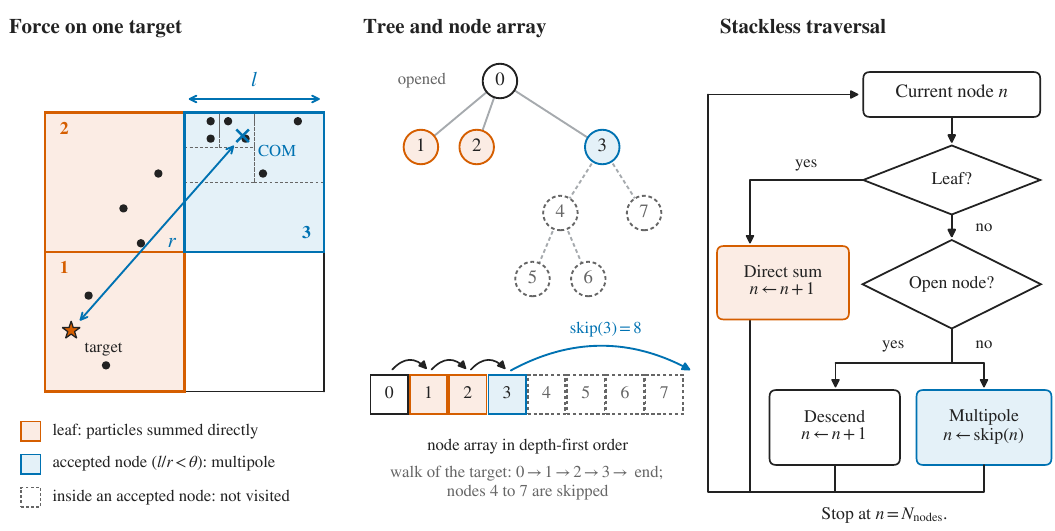}
    \caption{Tree walk for the gravitational force on one target particle,
    illustrated in two dimensions with the particles and the tree of
    \cref{fig:nori_tree_construction}, in which every cell with more than
    three particles is subdivided ($K=3$ for illustration; the code default is $K=8$).
    Left: the cells used by a walk with the geometric opening criterion
    ($\theta=0.7$). The particles of the two leaves (orange) are summed directly.
    For the node in the upper right (blue) we accept the multipole expansion, because its side length $l$
    is smaller than $\theta$ times the distance $r$ between the target and its
    centre of mass (COM). The
    cells inside it (dashed) are not visited.
    Middle: the tree and its depth-first node array with the path of the walk.
    Opening the root and summing a leaf both lead to the next entry of the
    array, whereas accepting a node leads to its skip pointer (e.g. $\mathrm{skip}(3)=8$), the first entry after its subtree.
    Right: the traversal logic for a node $n$; the walk ends when $n$ reaches
    the number of nodes, $N_{\rm nodes}$. }
    \label{fig:bh_tree}
\end{figure*}

The total gravitational force on a given target particle $i$ is calculated by a tree walk starting at the root of the tree, illustrated in two dimensions in \cref{fig:bh_tree}.
For each internal node, we test an opening criterion. If the node is accepted, we use its multipole expansion; otherwise, we open it and test its daughter nodes. Particles in a leaf are summed directly.
The nodes can be stored in a depth-first node array, in which every node is immediately
followed by the nodes of its subtree. Each node also stores a skip
pointer, the index of the first array entry after its subtree. The walk
then needs no stack: opening a node continues with the next array
entry, whereas accepting a node or summing a leaf continues at
its skip pointer.
A standard geometric opening criterion accepts a cell
of side length $l$, whose centre of mass lies at distance $r$ from the target, if
\begin{equation}
    l < \theta r,
    \label{eq:bh_criterion}
\end{equation}
where the opening angle $\theta$ controls the accuracy. Alternatively, the acceleration-based
criterion of the \textsc{gadget} codes \citep{springel2001gadget,Springel2005Gadget2,springel2021simulating} accepts a cell if
\begin{equation}
    \frac{GM}{r^2} \left(\frac{l}{r}\right)^p
    \leq
    \alpha\,|\bm{a}_{\rm old}|,
    \label{eq:rel_criterion}
\end{equation}
where $M$ is the cell mass, $\alpha$ a dimensionless tolerance, and $\bm{a}_{\rm old}$ the acceleration of
the target particle in the previous time-step. The exponent $p$ is the order of the leading term omitted from the multipole expansion: $p=2$ for monopoles, since the dipole vanishes about the centre of mass, and $p=3$ when quadrupoles are included. Following \textsc{gadget-4} \citep{springel2021simulating}, we refer to these two choices as O1 and O3, after the highest multipole order included: 1 for the monopole and 3 for the quadrupole.
The acceleration-based criterion of \cref{eq:rel_criterion} spends more computational work
in regions where large force contributions cancel and the resulting
acceleration is small.
The average cost per particle for the full tree walk is approximately proportional to the depth of the tree, i.e.\ to $\log N$ for a homogenous particle distribution, so the total cost for $N$ particles scales as $\mathcal{O}(N\log N)$. A one-sided walk does not, however, evaluate
the interaction between two particles or cells symmetrically. The force on
particle $i$ from a cell containing particle $j$ is generally not paired with
the opposite force. Linear momentum is therefore only conserved to the
accuracy of the multipole approximation.
Symmetric fast multipole methods (FMMs) apply cell--cell
interactions to both sides and conserve momentum to machine precision
\citep{Dehnen2000,Dehnen2002,springel2021simulating}.

\subsection{SIDM}
\label{subsec:sidm_methods}
Most particle-based SIDM implementations approximate the collision operator $\mathcal{C}$
by random scattering events between nearby simulation particles. In this paper, we restrict
ourselves to equal particle masses and elastic, isotropic interactions, while allowing the
cross-section per unit mass, $\sigma/m$, to depend on the relative velocity.
For an ordered particle pair $(i,j)$, i.e.\ the pair as seen from particle $i$, we estimate the contribution of
particle $j$ to the scattering rate of particle $i$ as
\begin{equation}
    \hat{\Gamma}(i|j)
    =
    m_j\,W(r_{ij};h_i)\,
    \frac{\sigma(v_{ij})}{m}\,v_{ij},
    \label{eq:rate}
\end{equation}
where $r_{ij}=|\bm{x}_i-\bm{x}_j|$ and $v_{ij}=|\bm{v}_i-\bm{v}_j|$ are the separation and relative speed of the pair, and $W$ is the scattering kernel. The smoothing length $h_i$ is
chosen such that the compact support of $W$ contains a specified number of neighbours. Because the estimate uses the smoothing length of particle $i$, $\hat{\Gamma}(i|j)$ and $\hat{\Gamma}(j|i)$ generally differ; the reverse pair $(j,i)$ is treated below. We use
the cubic-spline kernel
\begin{equation}
    W(r;h) = \frac{8}{\pi h^3}
    \begin{cases}
        1-6u^2+6u^3, & 0\leq u\leq \tfrac{1}{2},\\[2pt]
        2(1-u)^3, & \tfrac{1}{2}<u\leq1,\\[2pt]
        0, & u>1,
    \end{cases}
    \qquad u=\frac{r}{h},
    \label{eq:kernel}
\end{equation}
which is normalized to unity and has compact support for $r\leq h$. Other
common choices include a top-hat kernel \citep{Robertson2017SIDM} or a symmetrized rate estimator based on the overlap integral of two cubic splines \citep{rocha2013cosmological}.
In \citet{ZierSidm2026}, we compared the cubic-spline and overlap estimators and found no noticeable difference between them in simulations of gravothermal collapse.
The smoothing length can be determined iteratively by choosing a trial radius $h$ and finding all neighbours within it.
An efficient neighbour search can use the gravity tree \citep[e.g.][]{rocha2013cosmological} or a dedicated tree optimized for neighbour searches \citep{ZierSidm2026}.

The probability that particle $i$ scatters with a neighbour $j$ found during its neighbour search is
\begin{equation}
    P(i|j)
    =
    \frac{1}{2}\,\hat{\Gamma}(i|j)\,\Delta t,
    \label{eq:probabilityScatter}
\end{equation}
where $\Delta t$ is the time-step and the factor $1/2$ accounts for the reverse trial $(j,i)$, which is evaluated separately when $j$ finds $i$ in its own search, so that the unordered pair receives the symmetrized rate $\tfrac{1}{2}[\hat{\Gamma}(i|j)+\hat{\Gamma}(j|i)]$.
In practice, a uniform random number $x\in[0,1)$ is drawn, and the pair scatters if $x<P(i|j)$.
For
equal particle masses, an accepted elastic scattering event changes the pair
velocities to
\begin{equation}
    \bm{v}'_i =
    \bm{v}_{\rm com}+\frac{v_{ij}}{2}\,\hat{\bm{e}},
    \qquad
    \bm{v}'_j =
    \bm{v}_{\rm com}-\frac{v_{ij}}{2}\,\hat{\bm{e}},
    \label{eq:sidm_kick}
\end{equation}
where $\bm{v}_{\rm com}=(\bm{v}_i+\bm{v}_j)/2$ is the
centre-of-momentum velocity and the direction
$\hat{\bm{e}}$ is sampled uniformly on the unit sphere, independent of the incoming direction, since the scattering is isotropic. This update
conserves the momentum and kinetic energy of the pair to round-off.

This scattering scheme has three important numerical limitations.
First, the two simulation particles interact at a finite separation. The resulting
kick is not central, so angular momentum is not conserved exactly. It is
recovered only in the limit $h\to0$ \citep{Fischer2024Energy}. Second, every scattering probability has to remain
small. Otherwise, multiple interactions involving the same particle introduce
corrections of order $P^2$. They can also violate energy conservation if all
events are applied using outdated velocities. Third, the scattering kernel must resolve the spatial scales relevant to heat transport.
A scattering kernel larger than the mean free path
$\lambda=[\rho\,(\sigma/m)]^{-1}$ can introduce excess heat transport, although the acceptable ratio depends on the system and requires convergence tests \citep{Fischer2025Collapse}.

\subsection{Time integration}
\label{sec:time_methods}
Between scattering events, \cref{eq:eom} can be integrated with a
kick--drift--kick leapfrog scheme, which consists of a first half-kick
of the velocities, a drift of the positions, and a second half-kick:
\begin{align}
    \bm{v}_i^{\,n+1/2} &=
    \bm{v}_i^{\,n}+\bm{a}_i^{\,n}\,\frac{\Delta t}{2}, \nonumber\\
    \bm{x}_i^{\,n+1} &=
    \bm{x}_i^{\,n}+\bm{v}_i^{\,n+1/2}\,\Delta t, \label{eq:kdk}\\
    \bm{v}_i^{\,n+1} &=
    \bm{v}_i^{\,n+1/2}+\bm{a}_i^{\,n+1}\,\frac{\Delta t}{2}. \nonumber
\end{align}
For a constant time-step and a force derived from a time-independent Hamiltonian, this method is second-order
accurate, time-reversible, and symplectic. Its energy error remains bounded
and oscillates instead of growing systematically
\citep{Quinn1997,HairerLubichWanner2006}. These properties no longer hold
exactly if the time-step changes, the force approximation differs between
steps, or scattering kicks are inserted.

A common choice for the gravitational time-step of particle $i$ is
\begin{equation}
    \Delta t_{{\rm grav},i}
    =
    \sqrt{\frac{2\eta\,\epsilon}{|\bm{a}_i|}},
    \label{eq:dtgrav}
\end{equation}
where $\eta$ controls the time integration accuracy
\citep{springel2001gadget,Power2003}.
A separate condition is needed for
self-interactions.
In \citet{ZierSidm2026}, we introduced a conservative criterion
\begin{equation}
    \Delta t_{{\rm SIDM},i}
    \leq
    \frac{2C_{\rm SIDM}}
    {m_{\rm p}\,W(0;h_i)\,
    \max_j\!\left[v_{ij}\,\sigma(v_{ij})/m\right]},
    \label{eq:sidm_timestep}
\end{equation}
which limits the probability of every individual particle interaction to
a small value $C_{\rm SIDM}<1$.
The factor of two accounts for the
one half in \cref{eq:probabilityScatter}.
This criterion requires that several events involving the same particle within one step are applied in order, which is difficult to achieve in parallel, as discussed in \cref{sec:parallel_methods}.

Implementations that allow at most
one interaction per particle and step instead need to limit the total scattering probability, for example through
\begin{equation}
    \Delta t_i\sum_j\hat{\Gamma}(i|j)\leq\kappa
    \label{eq:timeStepCriterionMark}
\end{equation}
where $\kappa\ll1$ is the largest allowed total scattering probability per step.
\citet{vogelsberger2019evaporating} use a similar criterion, estimating the
rate from the local density and velocity dispersion, with
$\kappa=2.5\times10^{-3}$. Such limits on the total probability are
considerably more restrictive than \cref{eq:sidm_timestep} and make simulations deep into gravothermal collapse more expensive.

The time-step criteria above can be applied using either a global time-step or individual time-steps. With a global time-step, all particles are advanced using the smallest time-step required by any particle in the system. With individual time-steps, each particle is instead advanced according to its local gravitational and scattering conditions. In practice, individual time-steps are commonly organized into a hierarchy of discrete time-step levels, typically separated by factors of two, so that particles assigned to the same level are advanced together. Such hierarchical block time-step schemes avoid repeated force calculations for the slowly evolving particles in the outer halo. They are therefore efficient for systems spanning a wide range of dynamical time-scales. Moving particles between time-step levels breaks formal time reversibility. In addition, a scattering event between an active particle, i.e. one on a level currently being advanced, and an inactive one updates the two partners at different times. Both effects can lead to systematic energy errors in simulations of SIDM core collapse \citep{Fischer2024Energy}.

A global time-step becomes computationally expensive once only a small, dense core requires short steps. It does, however, keep all particles and interactions synchronized. The common time-step can still change when the most restrictive particle changes, but it then changes for all particles at once.

\subsection{Parallelization}
\label{sec:parallel_methods}

Large simulations need to distribute both memory and computational work across compute nodes.
Each compute node typically has several multi-core CPUs and possibly several GPUs. All CPU cores of one compute node can access its main memory directly (shared memory), but data on different compute nodes must be exchanged explicitly over the network, usually with the Message Passing Interface (MPI). Each independent MPI process is called an MPI rank, and typically several ranks run on one compute node.
A GPU (the device) is attached to a CPU and its memory (the host) and each GPU has its own memory and runs thousands of GPU threads concurrently.
Most general-purpose codes start from a spatial domain decomposition, in which
each rank owns the particles in one part of the simulation volume and only the corresponding part of the gravity tree.
Gravitational interactions across domain boundaries then require
communication. A rank can import the remote tree nodes needed for its
local walks \citep{springel2021simulating}. Alternatively, it can send target particles to the ranks that
own the relevant tree branches and receive the partial forces
\citep{Springel2005Gadget2}. Both methods avoid storing
all source particles on every rank, but require communication during the
force calculation. If all source data fit into the memory of a compute node or a GPU, one can instead replicate them and distribute only the walks of the target
particles \citep[e.g.][]{plimpton1995fast,Wang2015NBODY6GPU}. No remote tree data are then needed during a walk, at the cost
of additional memory and a global exchange of the results at every step.

Self-interactions add another complication. One accepted event modifies two
particles, and these particles may be owned by different ranks. Moreover,
several accepted events during the same time-step can involve the same
particle. Applying all of these events using the old velocities does not
conserve kinetic energy, although each individual scattering is elastic. A
parallel implementation must therefore prevent such conflicts or apply all
events in a consistent global order.
In \citet{ZierSidm2026}, we applied events in a consistent order using a parallelization scheme similar to the pairwise MPI processing of \citet{Fischer2021}, which allows the less restrictive time-step criterion in \cref{eq:sidm_timestep}.
Other implementations avoid multiple scattering events of the same particle by limiting the total scattering probability, as in \arepoOne \citep{vogelsberger2019evaporating}. After selecting scattering partners, this implementation also checks globally for particles involved in more than one event and rejects all events involving any such particle, as described in \citet{ZierSidm2026}. These conflicts are rare for sufficiently short time-steps, but rejecting them artificially lowers the scattering rate relative to the expected rate.

In the \gizmo implementations studied by \citet{Shen2022dSIDM,palubski2024numerical}, multiple scattering events involving the same particle on different MPI ranks can instead proceed without including earlier velocity changes in subsequent events.  The resulting artificial energy gain can become particularly important during core collapse, as the scattering rate and velocity dispersion in the core rise \citep{palubski2024numerical}.
An individual pair of events that share a particle can raise or lower the energy, but the mean change is positive.

\subsection{Conservation of energy and momentum}
\label{sec:conservation}

For an isolated system of constant-mass particles undergoing elastic interactions, the exact equations of the softened $N$-body system conserve the total energy
\begin{equation}
E_\epsilon
=
\frac{1}{2}\sum_i m_i|\bm{v}_i|^2
-\frac{G}{2}\sum_i\sum_{j\ne i}
m_i m_j\varphi_\epsilon(r_{ij}),
\label{eq:discrete_softened_energy}
\end{equation}
where $\varphi_\epsilon$ is the softened pair potential of \cref{eq:softened_potential}, given in \cref{app:SofteningKernel}. They also conserve the total linear momentum,
$\bm{P}=\sum_i m_i\bm{v}_i$. In numerical simulations, however, neither quantity is necessarily conserved exactly.

The main sources of conservation errors are the approximate gravity force of
\cref{sec:gravity_methods}, the time integration of \cref{sec:time_methods},
and the parallel application of scattering events discussed in
\cref{sec:parallel_methods}. These contributions must be controlled
separately: conserving energy in each elastic scattering event does not ensure conservation
of the total energy when gravity and time integration remain approximate.

Momentum errors can produce an unphysical drift of the system's centre
of mass. Energy errors can alter the heat balance of the collapsing core
and therefore change the collapse time or its late evolution
\citep{Fischer2024Energy,Mace2024,Fischer2025Collapse}.

\section{The \nori code}
\label{sec:nori}
\begin{figure*}
    \centering
    \includegraphics[width=\textwidth]{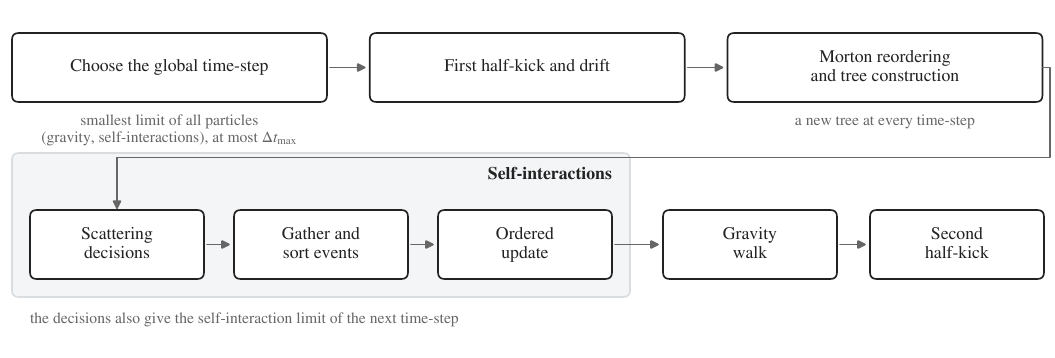}
    \caption{One global time-step in \nori. Top row: after the time-step
    has been chosen, the particles receive the first half-kick and are drifted.
    They are then sorted into Morton order, and the tree is rebuilt.
    Bottom row: on this tree, the self-interactions are computed first
    (scattering decisions, gathering and sorting of the accepted events, and
    the ordered update that applies them), followed by the gravity walk and the second half-kick.
    The scattering decisions also provide the self-interaction limit for the
    next time-step. CPUs and GPUs follow the same sequence. The exchanges of
    accelerations and smoothing lengths between replicas are omitted.}
    \label{fig:timestep}
\end{figure*}

\nori implements the gravitational and scattering methods of
\cref{sec:methods} for isolated, non-cosmological systems of equal-mass
particles. Self-interactions are assumed to be elastic and isotropic, with either a
constant or a tabulated velocity-dependent cross-section. All particles share one global time-step, as described in \cref{subsec:timestepAnatomy}.
\Cref{fig:timestep} gives an overview of the structure of one time-step.

We use the following vocabulary.
A \emph{replica} is a complete copy of the particle data that persists between
time-steps.
On CPUs, the MPI ranks on one \emph{shared-memory node} access a common
memory region through an MPI shared-memory window, sharing one replica
and one full copy of the tree.
On GPUs, the \emph{replicated GPU mode} gives every GPU a complete replica
and the full tree.
In the \emph{partitioned GPU mode}, the particles of one replica are instead
divided into $S$ equal \emph{segments}, each stored on a different GPU of the
same compute node. Every GPU still holds a full copy of the tree.
We call \emph{sources} the particles used to build the tree and \emph{targets} the particles for which we calculate gravitational forces or perform neighbour searches.

Both CPU and GPU paths follow the same execution model and share the
numerical routines for gravitational forces, scattering, and time integration.
Under the conditions described in \cref{subsec:binaryInvariance}, they
give bitwise-identical results, a property we refer to as \emph{binary invariance}.

\subsection{Tree construction}
\label{subsec:treeImplementation}

The tree is rebuilt from the particle positions at every time-step.
We use an octree whose leaf nodes contain up to $K$ particles. Allowing several particles per leaf regulates both the tree depth and the number of
nodes that need to be stored. It also changes the balance between direct
particle interactions and multipole evaluations,  because the walk never uses the multipole expansion of a leaf but sums the forces of its particles directly.
The most efficient choice
of $K$ depends on the relative costs of these operations and
of tree construction, as we show in \cref{subsec:impactBucketSize}.

In a serial implementation, such a tree can be constructed by inserting
particles one at a time and subdividing a cell whenever its occupancy
exceeds $K$. Parallelizing this procedure is more difficult, because
several ranks or GPU threads may attempt to modify the same branch. We instead
start from particles sorted along a Morton curve
\citep{WarrenSalmon1993,Bedorf2012,Karras2012}. This space-filling curve maps
three-dimensional positions onto a one-dimensional order in which nearby
particles are mostly close to each other. We then build the tree from the
bottom up, following \citet{Tokuue2024}: the leaves are determined from the
sorted keys, the internal nodes from the leaves, and the moments of a node
from those of its children. 
Compared to Peano--Hilbert ordering, Morton ordering simplifies key generation, particularly on GPUs, at the cost of somewhat
poorer spatial locality.

\begin{figure*}
    \centering
    \includegraphics[width=\textwidth]{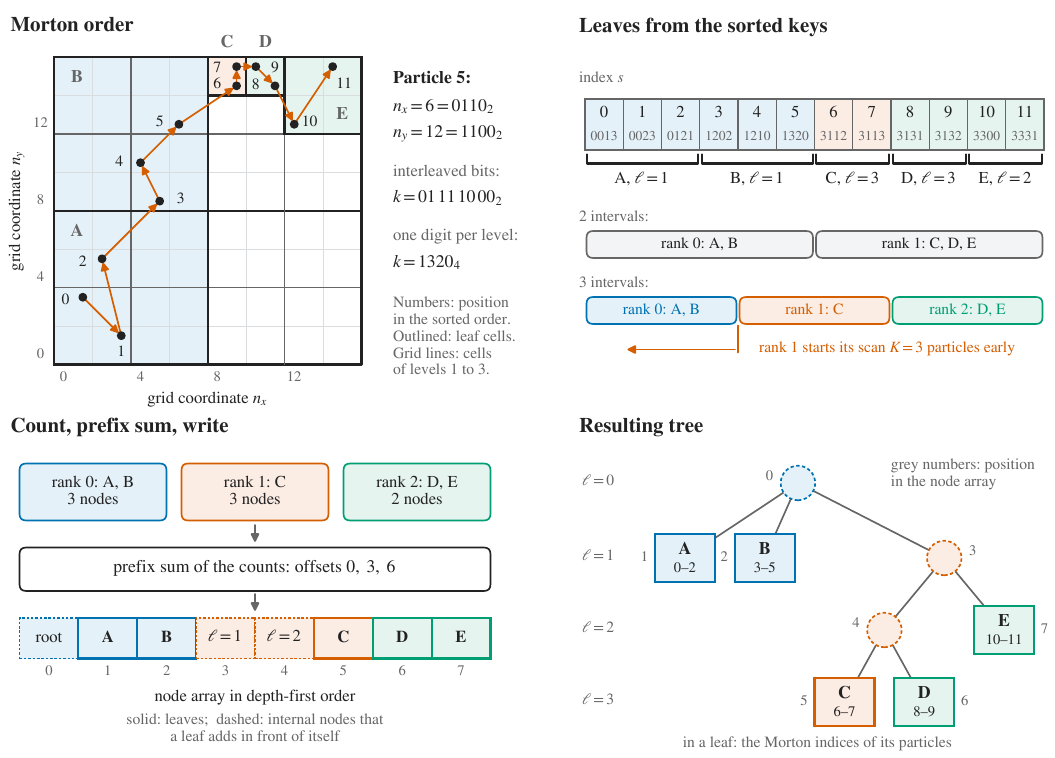}
    \caption{Construction of the tree from the sorted Morton keys,
    illustrated in two dimensions with twelve particles and at most $K=3$
    particles per leaf.
    Top left: the particles on a grid with 4 bits per coordinate. Interleaving the
    bits of the grid coordinates $n_x$ and $n_y$ gives the Morton key $k$ of a
    particle. Written with one digit per pair of bits, each digit of the key
    selects one of the four daughter cells at one level. The orange path
    connects the particles in the order of their keys, and the numbers are
    their Morton indices $s$. The outlined cells are the leaves A to E, and the
    grid lines are the cell boundaries of levels 1 to 3.
    Top right: the sorted particles with their keys. The particles of a leaf at
    level $\ell$ share the first $\ell$ digits of their keys.
    The leaves are the same whether the particles are divided into two or into
    three intervals.
    Bottom left: the count--scan--write allocation of the nodes for three
    intervals.
    Bottom right: the resulting tree, with the position of each node in the
    node array. The colours mark the rank that creates a leaf or
    node when the particles are divided into three intervals.}
    \label{fig:nori_tree_construction}
\end{figure*}

\subsubsection{Spatial ordering}

To compute the Morton keys, we first enclose the particle distribution
in a bounding cube with side length $L$ whose corner with the smallest
coordinates lies at the position $\bm{c}$, and map
the position $\bm{x}_i$ of each particle on to an integer grid of
spacing $\Delta = L/2^{42}$ within it,
\begin{equation}
    \bm{n}_i = \left\lfloor \frac{\bm{x}_i-\bm{c}}{\Delta} \right\rfloor ,
    \label{eq:nori_grid_coordinate}
\end{equation}
where the floor is taken for each component.
The Morton key of a particle is obtained by interleaving the bits of its
three integer coordinates: it begins with the most significant bit of
$n_x$, $n_y$, and $n_z$, followed by the next bit of each coordinate, and so on.
Every successive triplet of key bits then selects one of the eight
daughter cells of an octree cell, so that the first $3\ell$ bits identify
the cell at level $\ell$ that contains the particle. Sorting the particles
by their keys gives a one-dimensional ordering in which the
particles of any octree cell are contiguous, as the top left panel of
\cref{fig:nori_tree_construction} shows.
The 126-bit keys serve only to order the particles; the force and
scattering calculations continue to use the original double-precision
positions. The key resolution must be sufficient to distinguish all
particles, since the present construction requires strictly increasing
keys. The code stops if two particles fall into the same cell of the deepest level.

After every drift operation, we recompute the keys and physically reorder the
particle positions and velocities into Morton order before building the
tree. Spatially nearby particles then mostly occupy neighbouring array
locations, improving locality during the subsequent force calculations and neighbour
searches. On both CPUs and GPUs, this reordering acts directly on the
particle arrays of the replica. The Morton index of a particle, i.e.\ its
position in the Morton-sorted sequence, is therefore also its array index
within the replica. Smoothing lengths,
previous-acceleration magnitudes $a_{\rm old}$, walk costs required for load balancing, and particle IDs
are permuted together with the positions and velocities.  The acceleration
vectors are not permuted: they have already been used for the first
half-kick and are overwritten by the next gravity calculation. The
order then stays fixed until gravity,
scattering, and the second half-kick have been completed.

The bounding cube has its corner at the minimum particle coordinates and a side length slightly larger than the largest extent of the particle distribution. Optionally, the tree grid is displaced randomly relative to the particles at every step, as done in \textsc{gadget-4} to reduce temporal correlations of the force errors
\citep{springel2021simulating}. We implement this by doubling the side length of the
bounding cube and shifting it by a random fraction of the original side length in each direction. This keeps all particles
inside the cube without changing their physical coordinates, at the cost
of one bit of key resolution. The
displacement is drawn from a random stream that depends on the time-step number, as described in \cref{subsec:binaryInvariance}, so every rank
constructs the same grid. All \nori simulations in this paper use this option, except the force-accuracy test of \cref{subsec:forceAccuracy}.

\subsubsection{Parallel construction}

The construction has four steps: we divide the sorted particles into leaves,
create the nodes, link them, and compute their multipole moments. In each
step, a node is determined by a few neighbouring keys or nodes alone, so that
different ranks or GPU threads can work on different parts of the tree
without coordinating with each other.

\textit{Leaves.} For two Morton keys $k_a$ and $k_b$, let $\mu(k_a,k_b)$
be the number of leading bit triplets they have in common. The two
particles then share a cell down to level $\mu$ and lie in different cells
at all deeper levels. We go through the particles in Morton order and start a
new leaf at the first particle that does not yet belong to a leaf. Let $s$ be the
Morton index of this particle. The cell of the new leaf has to satisfy two
conditions. It must not contain particle $s-1$, which belongs to the preceding
leaf, so its level has to exceed $\mu(k_{s-1},k_s)$. It must also not
contain particle $s+K$, as it would then hold more than $K$ particles, so
its level has to exceed $\mu(k_s,k_{s+K})$. We choose the largest cell that
satisfies both, i.e.\ the level
\begin{equation}
    \ell_s =
    \max\!\left[
        1,\,
        \mu(k_{s-1},k_s)+1,\,
        \mu(k_s,k_{s+K})+1
    \right],
    \label{eq:nori_leaf_level}
\end{equation}
where a comparison is omitted if particle $s-1$ or $s+K$ does not exist, and
the entry 1 ensures that the root itself is never a leaf. The leaf then
consists of particle $s$ and all following particles whose keys agree with
$k_s$ in the first $\ell_s$ triplets, and the next leaf starts at the first
particle after this group. The resulting leaves are those of an octree in
which every cell with more than $K$ particles is subdivided.

To find the leaves in parallel, we divide the sorted particles into contiguous
intervals and assign each interval to one rank (on CPUs) or one GPU thread.
A rank cannot simply apply the rule from the first particle of its interval
onwards, because this particle may lie in the middle of a leaf. However, a
leaf holds at most $K$ particles, so one of the $K$ particles before the
interval must be the first particle of a leaf. The rank therefore starts its
scan $K$ particles before its interval. The leaves it forms at first may be
too small, but none of them can extend beyond the end of the correct leaf it
lies in, so the scan arrives at a correct leaf boundary before it reaches the
interval. From there on, it reproduces the correct leaves, since
\cref{eq:nori_leaf_level} depends only on the keys. Each rank keeps only the
leaves whose first particle lies in its own interval, so that every leaf is
created exactly once, as illustrated in the top right panel of \cref{fig:nori_tree_construction}.

\textit{Nodes.} The internal nodes follow from the leaves. Consider again the
leaf that starts at particle $s$. All cells that contain particle $s$ at
levels up to $\mu(k_{s-1},k_s)$ also contain particle $s-1$ and have
therefore already been created together with an earlier leaf. The leaf thus
adds the cells at levels $\mu(k_{s-1},k_s)+1$ to $\ell_s-1$ as new internal
nodes, followed by the leaf itself; the first leaf also adds the root. If we
write down these nodes leaf after leaf, we obtain the depth-first node array
of \cref{sec:gravity_methods}, in which every node is directly followed by its
first child.

The number of nodes that a leaf adds is known from the keys before any node is
written. We use this to allocate the nodes with a count--scan--write pattern, shown in the
bottom left panel of \cref{fig:nori_tree_construction}. In a first pass,
every rank only counts the nodes of its interval. An exclusive prefix sum of
these counts, i.e.\ for each interval the total of the counts of all preceding
intervals, gives the position in the node array at which the nodes of this
interval begin. In a second pass, every rank writes its nodes into its own
range of the array. The ranges do not overlap, so no two ranks ever write to
the same memory. At this stage, a node stores its level, the Morton index of
its first particle, and the geometric centre of its cell; a leaf also stores
its number of particles.

\textit{Links.} Next, we determine for each internal node its skip pointer, introduced in
\cref{sec:gravity_methods}, and its number of particles. The subtree of a
node consists of the nodes that follow it in the array and whose first
particle still lies in its cell. Since the index of the first particle never
decreases along the array, the end of the subtree can be found with a binary
search. This gives the skip pointer, and the first particle of the node following the
subtree marks the end of the particle range of the subtree. Every node is handled independently of all others, so this step is
again carried out in parallel.

\textit{Moments.} Finally, we compute the centre of mass of every node and,
for O3, the quadrupole moments. The
centre of mass of a leaf is a sum over its particles, taken in Morton order.
The moments of an internal node are sums over its children, taken in depth-first order, which
requires the children to be complete first. On CPUs, each rank goes backwards
through its part of the node array, so that it reaches the children before
their parents. Nodes whose subtree extends into the part of another rank are
completed afterwards by one rank per shared-memory node, as the bottom right
panel of \cref{fig:nori_tree_construction} shows. GPUs instead
process the internal nodes level by level, starting with the deepest level,
and treat all nodes of one level in parallel.

Every sum follows a fixed order, so the moments, like the leaves and the
nodes, are the same for any division of the work, a property we use in
\cref{subsec:binaryInvariance}.

\subsection{Gravity solver}
\label{subsec:gravityImplementation}

The gravitational force is calculated with a one-sided tree walk, using either monopoles alone (O1) or monopoles and quadrupoles (O3). For the O3 mode, each node stores the six independent components of the symmetric
second-moment tensor
\begin{equation}
    Q_{ab} =
    \sum_{j\in\mathrm{node}} m_{\rm p}\,
    (x_{j,a}-s_a)(x_{j,b}-s_b),
    \label{eq:nori_second_moment}
\end{equation}
where $\bm{s}$ is the centre of mass of the node and $a,b\in\{x,y,z\}$ label the Cartesian components.
We keep the full tensor, including its trace. The trace does not contribute to the force for a Newtonian potential, but it does for a softened one. The force from a node is computed with the same spline-softened potential (see \cref{app:SofteningKernel}) as the force between two particles, so that the two are consistent inside the softening length. The quadrupole moments are stored in a separate array for the internal nodes only, since the walk never uses the moments of a leaf. Including them adds 48 bytes to the 64 bytes of an internal node and makes each accepted-node interaction more expensive, but reduces the number of nodes that
must be opened at a given force accuracy. The net effect on the cost of the walk is measured in \cref{subsec:impactBucketSize}.

For each target particle, we walk the tree from the root
and apply the opening criterion in \cref{eq:rel_criterion}. This uses
the acceleration from the preceding step to set the allowed error in
an individual interaction. The exponent is $p=2$ for O1 and $p=3$ for O3, and we choose the tolerance $\alpha$ separately for the two orders. In addition, two geometric safeguards open a node that lies too close to the target for its multipole expansion to converge, regardless of the error estimate of \cref{eq:rel_criterion}. The first is the geometric criterion of \cref{eq:bh_criterion} with the dimensionless $\theta = 1.2$: a node is opened if its side length $l$ exceeds $1.2\,r$, where $r$ is the distance from the target to its centre of mass. This value is deliberately loose, since the accuracy is controlled by $\alpha$, and only removes nodes that are large compared with their distance. The second refers to the geometric centre of the node rather than its centre of mass: a node is opened if all three coordinates of the target lie closer than $0.6\,l$ to those of the centre. This always opens a node that contains the target, even when its centre of mass is far from the target. Both safeguards are compile-time options, and all runs in this paper use the values given here. For the initial force calculation before the first time-step,
no preceding acceleration is available, and we instead use the
geometric criterion of \cref{eq:bh_criterion} with $\theta=0.7$.

 If a node is opened, the walk continues with the next entry of the depth-first array, which is its first child. If its multipole expansion is accepted instead, the walk jumps to the first entry after its subtree, which is stored in its skip pointer.
 When a leaf is reached, all its particles are summed
directly in Morton order. Every target particle can therefore be evaluated
independently, with a force summation order that is unaffected by the
rank or thread carrying out the walk. This is useful for both parallelization and reproducibility, but the force remains one-sided and does not conserve momentum exactly, as discussed in \cref{sec:gravity_methods}. The force errors and their effect on the evolution are assessed in \cref{sec:validation}. When needed for
diagnostics, the potential is accumulated in the same traversal,
with the finite spline self-potential subtracted analytically.

\subsection{Self-interactions}
\label{subsec:sidmImplementation}

We adopt the one-sided cubic-spline scattering estimator introduced
in \cref{subsec:sidm_methods}, as in our fiducial \arepoTwo implementation
\citep{ZierSidm2026}. Here we describe how \nori organizes the scattering calculation
in parallel while preserving consistent velocity updates for particles involved in several events.

\begin{figure*}
    \centering
    \includegraphics[width=\textwidth]{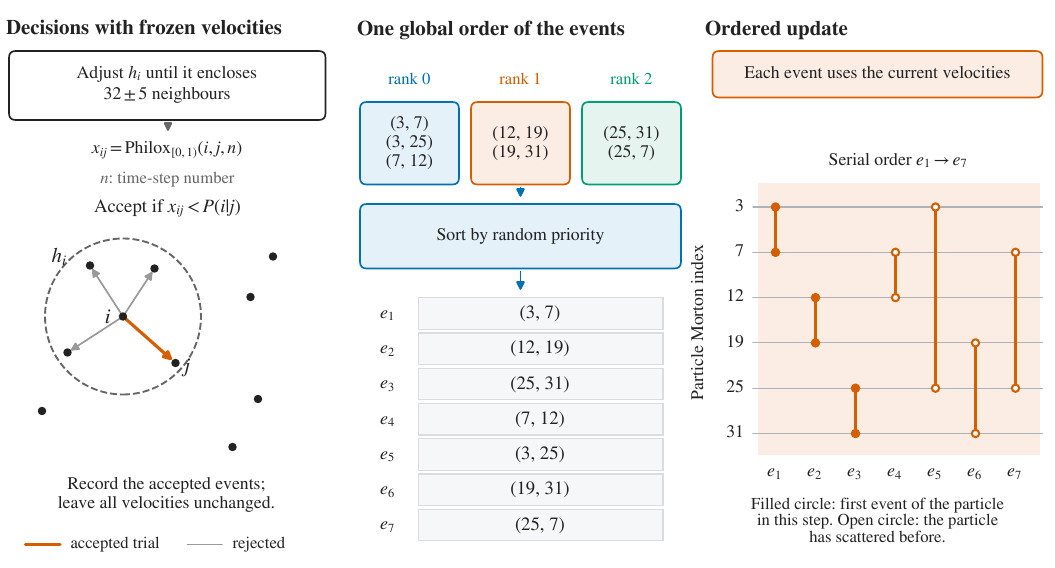}
    \caption{Parallel scattering decisions and ordered update of the velocities.
    Left: each particle $i$ adjusts its smoothing length $h_i$ to enclose
    the desired neighbour number and tests every neighbour $j$. The trial
    of the ordered pair $(i,j)$ is accepted if the uniform random number
    $x_{ij}\in[0,1)$, generated by the Philox generator from $i$, $j$, and the time-step number $n$ (\cref{subsec:binaryInvariance}), satisfies $x_{ij}<P(i|j)$. Accepted events are recorded, while all
    velocities are left unchanged (``frozen'').
    Middle: the accepted events of all ranks are gathered and sorted by a
    random 64-bit priority, with $i$ and $j$
    breaking ties. This gives one common order $e_1,\dots,e_7$.
    Right: the ordered update. One rank per replica applies the events serially in this order on
    the host. Each vertical bar connects the two partners of an event. An
    open circle marks a particle that already took part in an earlier event
    of this step. The event then uses its updated velocity.
    Particle indices are Morton indices.}
    \label{fig:sidm_replay}
\end{figure*}

In \arepoTwo, scattering events are decided and applied one after
another, so that each event sees the velocities left by the events before
it: pairs on the same MPI rank are processed in a local loop, and pairs
that span two ranks are processed afterwards by exchanging the current
velocities with one partner rank at a time. This order is set by the domain decomposition rather than by the times at which the events occur within the step, and it changes with the number of ranks. For \nori, we instead
separate the calculation of scattering probabilities from the
application of the accepted events. The probabilities are evaluated
in parallel from a common particle state, with velocities held fixed
until all decisions are complete. We then apply the accepted
scattering events one after another in a prescribed order, so that each event
uses the velocities left by the events before it. We call this second pass the
\emph{ordered update}, and the use of pre-scattering velocities
for all probabilities the \emph{frozen-rate approximation}. The scheme is illustrated in \cref{fig:sidm_replay}.
Its advantage is that only the accepted events require coordinated
velocity updates, while the much larger set of candidate pairs
can be tested independently.

\subsubsection{Neighbour search and scattering probabilities}

Since all particles are active in every time-step, the gravity tree contains the complete particle distribution at the time of the SIDM calculation. We reuse it for the neighbour searches. By comparison, \arepoTwo needs a dedicated tree because of its hierarchical gravity time integration \citep{ZierSidm2026}. For a trial value of the
smoothing length $h_i$, we discard every tree node whose
cell does not intersect the sphere of radius $h_i$ around particle $i$ and test the particles
in the remaining leaves individually.
A search stops as soon as
the number of neighbours exceeds the upper end of the allowed range, in which case $h_i$ is reduced.
We enlarge or shrink the radius by
a factor of $1.26\approx2^{1/3}$, switching to bisection in $h_i^3$ once
the target neighbour count is bracketed.
On CPUs, we
iterate each particle to convergence before moving on to the next one.
On GPUs, each kernel launch instead performs one iteration for all
particles that have not yet converged. During the launch, every particle
whose smoothing length is not accepted appends itself to a compact list,
which is the input of the next launch, so that the work per launch
shrinks with the number of unconverged particles. The scattering trials are
evaluated in the iteration that accepts the smoothing length, using the
neighbour list of that search.

We use a desired neighbour number of $32\pm5$, not counting particle $i$
itself. The search radius is capped at a maximum smoothing length, a
compile-time constant, so that it cannot grow without bound around
isolated or escaping particles. A particle whose search reaches this
maximum with too few neighbours is accepted as it is. Each search starts
from a prediction made at the end of the previous step. Assuming that the
neighbour number scales as $h_i^3$, we predict
$h_i-(N_i-N_{\rm neigh})\,h_i/(3N_i)$ from the accepted smoothing length
$h_i$ and its neighbour number $N_i$, where $N_{\rm neigh}$ is the desired
neighbour number. A particle without neighbours uses $N_i=1$ in the
denominator, and the prediction is capped at the maximum smoothing length
as well.
During the iteration, we keep the largest smoothing length that gave too few neighbours and the smallest one that gave too many. If these two bounds agree to a relative difference of $10^{-6}$ and no admissible smoothing length has been found between them, we accept the lower bound, for which the particle has fewer neighbours than desired. These cases still use the complete neighbour
list for the accepted smoothing length.

For every neighbour within $h_i$, we evaluate
\cref{eq:probabilityScatter} for the trial of the ordered pair $(i,j)$
and compare the result with a uniform
random number. The reverse trial $(j,i)$ is evaluated separately
when particle $j$ finds $i$ in its own search, which accounts for the
factor of one half in the probability. For velocity-dependent
interactions, the cross-section is interpolated linearly in
relative speed between the entries of a table with increasing
speeds. Below the tabulated range, we use the cross-section of the first entry; a relative speed above the range terminates the
calculation. Each accepted event is stored in 128 bits: the two 32-bit Morton indices of its partners and a 64-bit random priority used to sort the events, as described in \cref{subsec:binaryInvariance}. The amount of event data to be communicated scales with the number of accepted events. Since the particles are not reordered between the scattering decisions and the ordered update, the stored indices still refer to the same particles, and the ordered update uses them directly to read and update the velocities in the particle arrays of the replica.

\subsubsection{Consistent multiple scattering}

We gather the accepted events from all MPI ranks and sort them on
the host by their 64-bit priority, using the two particle
indices to break any ties. This gives the same total order regardless
of the sequence in which different ranks or GPU threads found or communicated their
events. The priority is drawn together with the acceptance number of the trial, as described in \cref{subsec:binaryInvariance}, so the position of an event in the list is unrelated to the probability with which it was accepted. For each event, we recompute the
centre-of-momentum velocity and relative speed from the velocities
currently stored for the two partners, and apply the isotropic update of
\cref{eq:sidm_kick}. The direction $\hat{\bm{e}}$ is sampled
by the rejection method of \citet{marsaglia1972choosing} from a separate random number stream, which avoids the trigonometric functions used in \arepoTwo, whose rounding differs between math libraries. Each
scattering event then conserves the momentum
and kinetic energy of its pair to floating-point round-off, even when a
particle has already scattered earlier in the list. In
particular, there is no need to cancel an accepted event merely
because one of its particles occurs in another event.

Every rank receives the complete event list, and one MPI rank per replica applies it serially in this order as shown in \cref{fig:sidm_replay}. On CPUs, this rank updates the shared velocity array of its shared-memory node directly. On GPUs, it copies the velocities of all particles involved
in accepted events from the GPU memory to the host, each particle once, applies the events, and copies the updated velocities back. No other particle data are copied to the host, so the size of this transfer is set by the number of accepted events and not by $N$. Since every replica applies the same list, no subsequent velocity exchange is required.

Events involving disjoint particle pairs commute, so the list could also be applied in parallel in groups of events without a common particle. In tests, applying the events in such groups in parallel on the GPU gave only a small gain: the global time-step is set by the particles with the highest scattering rates, so only a small fraction of all particles scatter in one step, and the ordered update takes a small fraction of the run time. We therefore apply the list serially on all backends. Together with sorting the event list and, on GPUs, gathering the velocities of the particles involved, it is the only serial part of the SIDM calculation, and its cost grows with the number of accepted events per step.

The ordered update conserves the momentum and energy of every event. Evaluating all
probabilities before any velocity is changed is nevertheless an approximation,
since a scattered particle's relative speeds to its other neighbours, and
hence the probabilities of its remaining trials, have changed. The sequential
processing of \arepoTwo \citep{ZierSidm2026} is an approximation of the same
order. Within a step, the events of a particle occur at random times, so a
later trial should see the updated velocity only for part of the step. The
frozen-rate approximation never includes this update, whereas sequential
processing always includes it for the events that follow in its processing
order. In both cases, the error is of second order in the pair probabilities
per step and vanishes as $\Delta t\to0$. We test the approximation indirectly
in \cref{subsec:coreCollapse} by comparison with \arepoTwo.

\subsection{Time integration}
\label{subsec:timestepAnatomy}

The particles are advanced with a kick--drift--kick leapfrog method,
using a common time-step set by the most restrictive gravitational
or SIDM condition,
\begin{equation}
    \Delta t_{\rm lim} =
    \min\!\left[
        \Delta t_{\max},\,
        \min_i\Delta t_{{\rm grav},i},\,
        \min_i\Delta t_{{\rm SIDM},i}
    \right],
    \label{eq:nori_global_timestep}
\end{equation}
where $\Delta t_{\max}$ is a run-time parameter and the individual limits are given by
\cref{eq:dtgrav,eq:sidm_timestep}. On an integer timeline, which divides the simulated time interval into $2^{60}$ units, so that times and steps fit comfortably into 64-bit integers, we then take the largest power-of-two step that satisfies this limit and does not overshoot the final time.
The global time-step
keeps the two partners of every scattering event synchronized and avoids
a changing hierarchy of active and inactive particles. The cost of this choice once a small core sets the step for the whole halo is quantified in \cref{subsec:performanceFullSimulation} and discussed in \cref{sec:limitations}.

Starting from the current acceleration, we apply the first
half-kick and drift all particles over the full step. The
drifted positions and half-kicked velocities provide the state
for the SIDM decisions. We then sort the particles by their
new Morton keys and rebuild the tree. On this tree, we first carry out the scattering and then the gravity calculation, and apply the second half-kick with the new
acceleration. Since scattering needs no accelerations, its temporary storage uses the memory that subsequently holds them. Scattering changes only the velocities, so the tree remains valid for the gravity calculation that follows. All accepted events are applied within the same time-step.

For each particle, we estimate the next SIDM time-step limit from the largest
$v_{ij}\sigma(v_{ij})/m$ encountered in its own search and the
predicted smoothing length. If the local conditions change rapidly, the
actual pair probabilities can consequently exceed the intended
limit $C_{\rm SIDM}$. Such a step continues unchanged, but the code prints a warning, and a probability above unity
terminates the calculation before the ordered update. No SIDM limit is available
for the first step after initialization. This step is controlled by gravity and $\Delta t_{\max}$, which therefore has to be small enough to keep the pair probabilities of the first step below $C_{\rm SIDM}$.

\subsection{Parallelization}
\label{subsec:parallelization}

Because the particles and the tree are replicated, the complete source distribution has to fit into one compute node or, in the GPU modes, into one GPU or a group of $S$ GPUs. In return, the walks need no MPI communication, and we balance their cost by redistributing only the target particles.

\begin{figure*}
    \centering
    \includegraphics[width=\textwidth]{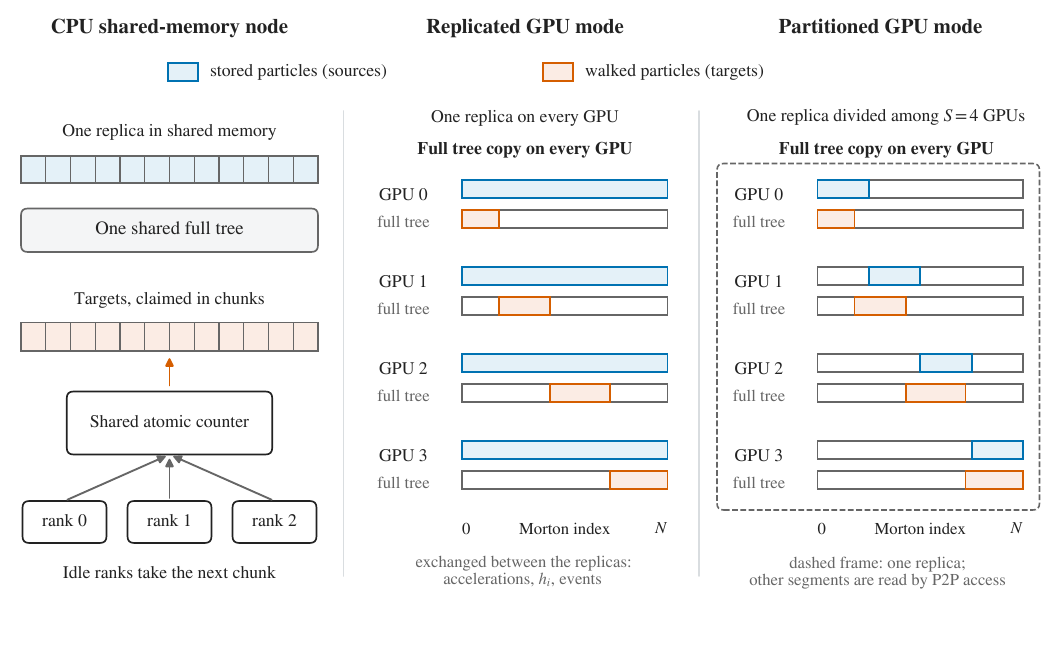}
    \caption{Storage of the particles and assignment of the work in the
    three parallel configurations. Blue marks the particles that a rank or GPU
    stores, orange the particles whose walks it
    carries out (the targets). The bars of the two GPU panels span the
    Morton-ordered particles.
    Left: on CPUs, the ranks of a shared-memory node share one replica and one
    tree. A rank that has finished a chunk of targets takes the next one by
    incrementing a shared counter.
    Middle: in the replicated GPU mode, every GPU stores a complete replica
    and walks one interval of targets.
    Right: in the partitioned GPU mode, the GPUs of a compute node share one
    replica (dashed frame). Each GPU stores one segment and reads particles of
    other segments through peer-to-peer (P2P) access.
    In both GPU modes, every GPU holds a complete tree.}
    \label{fig:parallel_layout}
\end{figure*}

\Cref{fig:parallel_layout} shows the three configurations defined in \cref{sec:nori}.
In all cases, the particles are stored in the current Morton
order.
Which rank or GPU carries out the walks of a particle does not depend on where the particle is stored.

\subsubsection{CPU parallelization}

The ranks of a shared-memory node divide the replica and the tree into contiguous intervals.
After each drift, every rank computes and sorts the keys of its
interval, and the sorted intervals are merged. The ranks then permute
the particle fields in place and cooperatively construct the shared
tree. The available memory is used for a single
large particle distribution, rather than for one copy of
the particles and the tree per CPU core.

Because the walk cost of a particle varies strongly across the halo, a static assignment of equal particle counts is unsuitable. Within a shared-memory node, ranks claim chunks of consecutive Morton indices from a shared atomic counter,
which each rank increments to claim the next chunk without conflicts. The chunk size is chosen such that each rank claims about 32 chunks per walk, but limited to between 256 and 4096 particles.
This dynamic scheduling distributes expensive regions among the
available cores while retaining spatial locality. It is used for both the gravity walks and the SIDM neighbour searches. With several
shared-memory nodes, the target particles are first divided between
them with the deterministic cost-weighted partition described for
GPUs below. After the walks, the shared-memory nodes exchange the
accelerations and smoothing lengths of their targets.

\subsubsection{Load balancing and GPU parallelization}

In replicated GPU mode, one MPI rank controls each GPU. The replica stays in the memory of the GPU for the whole simulation and is not copied back to the host in full between time-steps.
 Only memory management, communication, and scheduling differ from the CPU implementation. The GPU code is written against a thin abstraction layer, in which every kernel is a function that is applied to a range of indices. This layer compiles directly to CUDA or HIP, the GPU programming interfaces of NVIDIA and AMD, respectively, or to a serial loop on the CPU, which serves as a debug backend.

Each target particle is handled by one GPU thread. Each thread independently traverses the depth-first node array using the skip pointers of \cref{sec:gravity_methods} and reads nodes and particles directly from global memory, without shared-memory staging or interaction lists shared within a warp, the group of 32 (NVIDIA) or 64 (AMD) threads that executes the same instruction at the same time. With the relative opening criterion, the nodes accepted by neighbouring targets differ, so warp-level group walks offer little benefit, and the Morton order of the targets provides most of the locality. The neighbour search proceeds in rounds, each compacting the particles whose smoothing length has not converged into a new list, so that from the second round onwards the threads no longer follow the Morton order. Within the leaf loops, only a small fraction of the threads of a warp is typically active, because the threads reach leaves at different steps of their walks and the others wait. Node interactions, in contrast, keep most of them busy. Direct particle interactions are therefore less efficient on GPUs than on CPUs, which is why the optimal leaf size is smaller on GPUs, as shown in \cref{subsec:impactBucketSize}.

The particles are reordered with stable radix sorts.
We rebalance the work after every reordering. As the cost of a particle, we
use the number of nodes and particles that its gravity walk visited in the
previous step. This walk cost is known only to the replica that carried out
the walk, and exchanging it for all particles would be expensive. Each replica
therefore adds up the costs of its particles in bins of about 1024 consecutive
Morton indices, and only this small table is summed over all ranks. Every rank
then holds the same table and cuts the Morton sequence, at bin boundaries,
into one contiguous interval of targets per rank with approximately equal
total cost. The sums use integer arithmetic, so that the boundaries are
reproducible. We also require that no interval is empty and that none contains
more than twice the average number of particles, which bounds the temporary
memory of each rank. This
scheme balances the cost of the gravity walks. The neighbour searches use the
same intervals and are not necessarily balanced equally well.

After the gravity walks, each GPU sends the accelerations of its targets to
the other replicas, so that every replica holds the accelerations of all
particles before the second half-kick. The smoothing lengths are exchanged in
the same way after the neighbour searches. Between compute nodes,
the data are sent either by a GPU-aware MPI library, which reads GPU memory
directly, or through page-locked host memory. Within a compute node, GPUs
with peer-to-peer (P2P) access write the data directly into each other's arrays. The accepted
scattering events always pass through the host. In contrast to the walks, sorting, tree construction, and the ordered update are repeated on every replica, so that their share of the run time grows when a fixed problem is distributed over more GPUs.

\subsubsection{Partitioned GPU mode}

The partitioned GPU mode raises the memory limit of the replicated mode on
compute nodes whose GPUs have fast P2P access. Each of the $S$ GPUs of a compute node stores one segment, an equal share of the Morton-ordered particles of the replica.
When a walk
needs a particle of another segment, it reads it through P2P access from the
GPU that stores it.

Since the segments follow the Morton order, the particles have to be
redistributed among the GPUs after every drift. Each GPU first sorts the keys
of its own particles. We then determine the keys that cut the Morton sequence
into segments of equal size. Each GPU sends its keys to their new segments,
where they are sorted again, and the particle data follow in the same order.

The targets are divided by cost as in the replicated mode, one interval per
GPU, and the intervals are assigned in the order of the segments. Since the
segments have equal size, the two sets of boundaries do not coincide: a GPU
mostly walks particles of its own segment, plus some that are stored on a
neighbouring GPU, as the right panel of \cref{fig:parallel_layout} shows. The tree is built jointly, in two stages. In the first stage, each GPU finds the leaves among the particles of its own segment and creates their nodes, including the centres of mass of the leaves. It writes these nodes into its own copy of the tree and into the copies of the other GPUs of the replica, so that every copy then contains all nodes. In the second stage, every GPU determines the links and the moments of the internal nodes for the whole tree in its own copy. This work is repeated on every GPU, but it requires no further exchange of nodes. The ordered update is unchanged, except that the rank of the first GPU of the replica carries it out and reads and writes the velocities on all GPUs of the replica.

Partitioning reduces the memory for the particles on each GPU by a factor of
$S$. The tree and the temporary buffers are not reduced, and they now
fix the maximum size of a simulation.

\subsection{Binary invariance across parallel configurations}
\label{subsec:binaryInvariance}
Binary invariance, i.e.\ bitwise-identical results independent of the numbers of ranks and GPUs, the number of segments $S$, and, by construction, the hardware, allows the GPU implementation to be debugged on CPUs and makes every run exactly reproducible.
Floating-point addition is not associative because of round-off, so binary invariance requires the same order of operations.
Random events must also receive the same random numbers independently of the setup.

In \nori, the Morton ordering and tree construction already provide
the same source sequence, node geometry, and moment sums
independently of the parallel distribution.
Each target then
traverses this tree in the same order, so dynamic work
assignment does not change its force accumulation.
For the
random number generation, we use the counter-based Philox4x32-10
generator \citep{Salmon2011Philox}, which computes its output as a fixed function of two inputs, a counter and a key, instead of from an evolving internal state.
For the trial of an ordered pair, the counter is formed from the current Morton indices of its two particles and the time-step number.

We use three separate random streams, which differ by the generator key,
to decide whether a pair scatters, to choose the scattering direction,
and to shift the tree grid.
Each call of the generator returns four 32-bit words. For a scattering
trial, two of them give the uniform number for the acceptance test,
and the other two give the 64-bit priority of
the resulting event. Each interaction therefore
receives the same random numbers, regardless of which
rank or GPU thread evaluates it or the order in which interactions
are visited.

 Finally, the CPU and GPU implementations have to round every floating-point operation in the same way. Both use the same source code for the arithmetic, and every quantity is stored and computed with the same precision on all backends.

Fused multiply-add (FMA) operations pose a particular difficulty. An FMA operation evaluates $ab+c$ with a single rounding at the end, whereas a separate multiplication and addition round twice, so that the two results can differ. Compilers are allowed to merge a multiplication and a following addition into an FMA operation, and they decide differently for CPUs and GPUs and from one compiler version to the next. We therefore switch off this automatic contraction for the whole code. Where an FMA operation is useful, as in the force and scattering kernels, we write it explicitly: a requested FMA is exactly defined and gives the same bits on every platform. For the same reason, we disable all compiler optimizations that reorder or simplify floating-point expressions, such as those of the fast-math options. The build system checks the compiler options recorded in every object file and stops if one of them was compiled with contraction or with such an optimization.

Mathematical library functions are a further source of differences, because their rounding is not standardized and differs between libraries and between CPUs and GPUs. The arithmetic that enters the simulation therefore uses only additions, multiplications, divisions, square roots, and FMAs, whose results are exactly defined. Other functions are replaced: the cube root in the smoothing-length iteration is computed with a fixed number of Newton steps, and the scattering direction is drawn with a rejection method instead of trigonometric functions. Lastly, a quantity that several replicas need is always copied from the replica that computed it and never summed over ranks in floating point, since such a sum would depend on the number of ranks.

With these choices, the evolution is bitwise identical for fixed initial conditions and numerical parameters across CPU and CUDA runs with different parallel configurations. We verified this for the simulations presented in the rest of the paper by comparing snapshots of CPU and GPU runs with different numbers of ranks, GPUs, and segments.

\subsection{Input, output, and diagnostics}
\label{subsec:IO}
\nori reads and writes HDF5 snapshots in the
\textsc{gadget}/\arepo snapshot format 3, so that existing initial conditions and analysis tools can be reused.
The particle mass, the softening length, a constant cross-section, if one is used, and most other numerical parameters are set at compile time, which allows additional compiler optimizations that are useful on GPUs. A velocity-dependent cross-section is instead read at run time from the table described in \cref{subsec:sidmImplementation}.
The code requires dark-matter-only initial conditions whose mass table entry matches the compiled particle mass.
The particle IDs have to be a permutation of the integers from one to $N$, in any order.
Snapshots store the particles in their Morton order at the time of output, independently of the parallel configuration.
The binary restart files written by \nori allow a bitwise-identical continuation with the same number of ranks.

During the simulation, \nori records the kinetic, potential, and total
energy, the total linear momentum, and the total force $\sum_i m_i\bm{a}_i$, which measures the momentum error of the one-sided gravity calculation. For the energy comparisons in \cref{sec:validation}, the potential energy is instead recomputed by direct summation in post-processing, as discussed in \cref{app:ImpactEnergyMeasurement}.
The code also records the timings of its different phases and the peak usage of its memory arena.

\section{Validation}
\label{sec:validation}
We test \nori on idealized problems, first for gravity alone and then with self-interactions.
We compare with direct summation and analytic profiles where available, and otherwise with our SIDM implementation in \arepoTwo \citep{ZierSidm2026}.
We cannot repeat the periodic-box thermalization test of \citet{ZierSidm2026}, because \nori does not support periodic boundary conditions.

\subsection{Force accuracy}
\label{subsec:forceAccuracy}

The multipole expansions of the tree introduce a force error.
We measure it by comparing the tree force with direct summation for a static, isotropic Hernquist sphere \citep{hernquist1990analytical}, following the force test of \citet{springel2021simulating}.
We choose $G=1$, a total mass $M=1$, and a scale radius $a=1$, and draw $N=10^6$ equal-mass particles from the Hernquist profile within $r<100a$.
The Plummer-equivalent softening is $\epsilon=10^{-6}a$, far below the particle separations, so that the softening is negligible for this test.

For every particle, we calculate the direct-summation acceleration
$\bm{a}_{i,\mathrm{direct}}$ once and define the relative tree-force error as
\begin{equation}
    \delta a_i =
    \frac{\left| \bm{a}_{i,\mathrm{tree}}
    -\bm{a}_{i,\mathrm{direct}} \right|}
    {\left| \bm{a}_{i,\mathrm{direct}} \right|}.
    \label{eq:force_error}
\end{equation}
We summarize the error distribution by its 90th percentile, $\delta a_{90}$, the error that 90 per cent of the particles do not exceed.
As described in \cref{subsec:gravityImplementation}, the relative opening
criterion of \cref{eq:rel_criterion} needs the acceleration of the previous
step, which the first force calculation does not have. In this test, we
use $|\bm{a}_{i,\mathrm{direct}}|$ instead, so that every tree
walk uses the relative criterion and the geometric opening of the initial force calculation
does not enter the comparison.

\begin{figure}
    \centering
    \includegraphics[width=\linewidth]{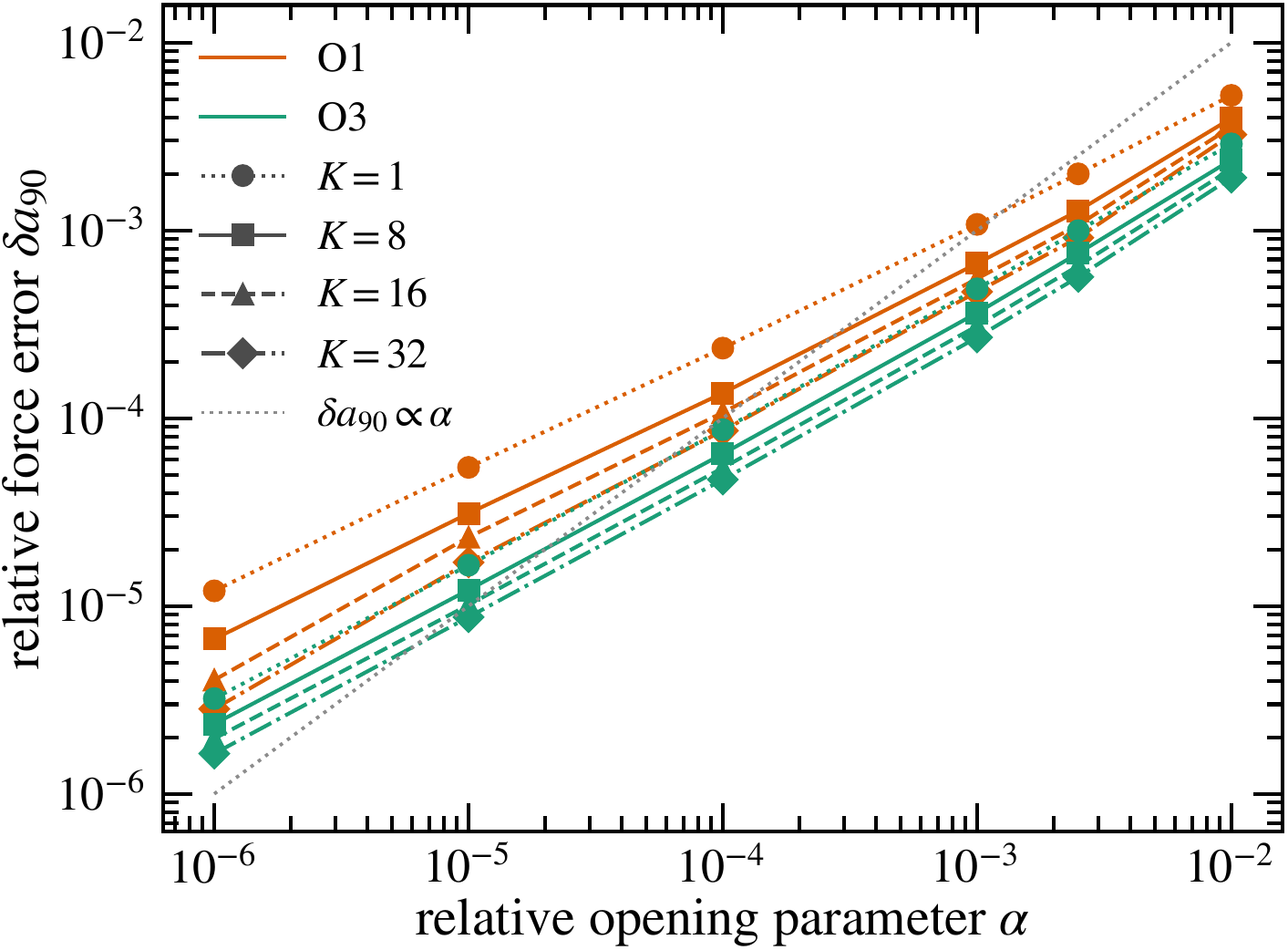}
    \includegraphics[width=\linewidth]{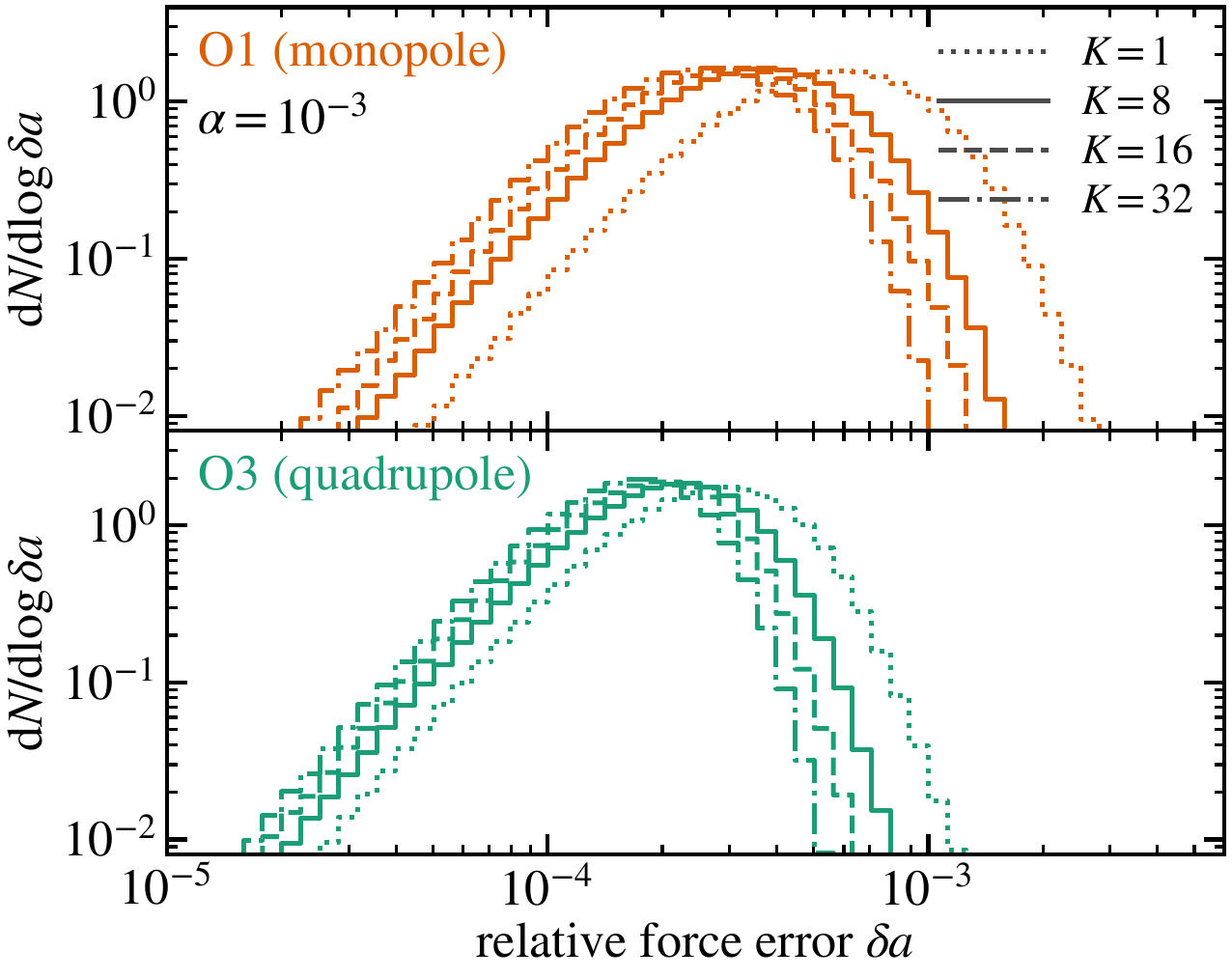}
    \caption{Accuracy of the tree force for an effectively unsoftened Hernquist sphere
    sampled with $N=10^6$ particles. Top: 90th-percentile relative error,
    $\delta a_{90}$, with respect to direct summation as a function of the
    opening parameter $\alpha$. Orange and green curves show monopole (O1)
    and quadrupole (O3) gravity; symbols and line styles distinguish the
    maximum numbers of particles per leaf, $K=1$, 8, 16, and 32. The grey
    dotted line shows $\delta a_{90} \propto \alpha$ for comparison. Bottom: normalized
    distributions of the relative error $\delta a$ at $\alpha=10^{-3}$ for
    the same four values of $K$, separately for O1 and O3.}
    \label{fig:forceError}
\end{figure}

We test all 48 combinations of the monopole (O1) and quadrupole (O3)
expansions, the opening parameters
$\alpha=\{10^{-6},10^{-5},10^{-4},10^{-3},2.5\times10^{-3},10^{-2}\}$,
and the maximum numbers of particles per leaf $K=\{1,8,16,32\}$. \Cref{fig:forceError} shows that the error decreases monotonically as $\alpha$ is reduced, roughly as $\alpha^{0.7}$. A larger $K$ also reduces the error at fixed $\alpha$, including the high-error tail of the distribution, because a larger fraction of the force is evaluated by direct summation, which costs additional work. O3 reaches a given accuracy at a larger $\alpha$ than O1, i.e.\ with fewer opened nodes, but each node interaction costs more; \cref{subsec:impactBucketSize} compares the resulting cost.

We use these measurements to fix the opening parameters $\alpha$ used in the rest of the paper.
\arepoTwo uses O1 with $\alpha=0.0025$ and $K=1$. In this configuration, \nori has
$\delta a_{90}=2.0\times10^{-3}$, which we take as the error to match. For the fiducial $K=8$, O3 reaches approximately the same error at
$\alpha=0.0081$ and O1 at $\alpha=0.0045$. Unless stated otherwise, these are the opening parameters of all O3 and O1 runs in the rest of the paper. Only where \nori reproduces the \arepoTwo configuration in \cref{sec:performance} does it use O1 with $K=1$ and $\alpha=0.0025$.

\subsection{Collisionless equilibrium of an NFW halo}
\label{subsec:collisionlessEquilibrium}
To test the gravity solver and the time integration together over many
dynamical times, we evolve an NFW halo without self-interactions,
$\sigma/m = 0$, for $50\,\mathrm{Gyr}$. The halo starts in equilibrium, so
its radial profiles should not change.

We generate the initial conditions with
\textsc{SoftIsoICs}\footnote{\url{https://github.com/vinh-qtran/SoftIsoICs}}
\citep{Tran2024_1},
which samples an isotropic phase-space distribution function constructed in the softened potential \citep{Tran2025_3}, so that the halo is in equilibrium under the softened gravity of the
simulation. Within the virial radius $r_{200}$, we assume that the halo follows an NFW
profile \citep{navarro1997universal}. Beyond $r_{200}$, the density profile changes
to a modified exponential cut-off \citep{Springel1999},
\begin{align}
    \label{eq:nfw_inner_density}
    \rho_{r < r_{200}} (r) &= \frac{\rho_{\rm s}}{\left(r/r_{\rm s}\right) \left(1 + \left(r/r_{\rm s}\right)\right)^2}, \\
    \label{eq:nfw_outer_density}
    \rho_{r > r_{200}} (r) &= \frac{\rho_{\rm s}}{c \left(1 + c\right)^2} \left(\frac{r}{r_{200}}\right)^{\epsilon_{\rm d}} \exp\left(-\frac{r-r_{200}}{r_{\rm d}}\right).
\end{align}
We adopt a scale density $\rho_{\rm s}=1.10\times 10^{7}\,\mathrm{M_\odot\,kpc^{-3}}$
and a scale radius $r_{\rm s}=3.48\,\mathrm{kpc}$, which correspond to a virial mass
(the mass within $r_{200}$) of $M_{200}=10^{10}\,\mathrm{M_\odot}$ and a concentration
$c=13.08$. For simplicity, we set the decay scale to $r_{\rm d}=r_{200}$.
Continuity of the logarithmic density slope at $r_{200}$ then fixes the decay index $\epsilon_{\rm d}$,
\begin{equation}
    \label{eq:nfw_decay_exponential}
    \epsilon_{\rm d} = \frac{r_{200}}{r_{\rm d}} - \frac{1+3c}{1+c}.
\end{equation}
We sample $N_{200}=10^{7}$ particles within $r_{200}$ and extend the
sampling to $r_{\rm max}=3\,r_{200}$. The softening $\epsilon=0.016\,\mathrm{kpc}$
follows the criterion of \citet{Mace2024}, their equation~6, derived from \citet{vandenBosch2018} and evaluated for $N_{200}$ and $c$,
which they found to give converged collapse times in isolated SIDM haloes. This is the same halo with which we tested the accuracy
and performance of the \arepoTwo SIDM implementation in
\citet{ZierSidm2026}.

\begin{figure*}
    \centering
    \includegraphics[width=0.49\textwidth]{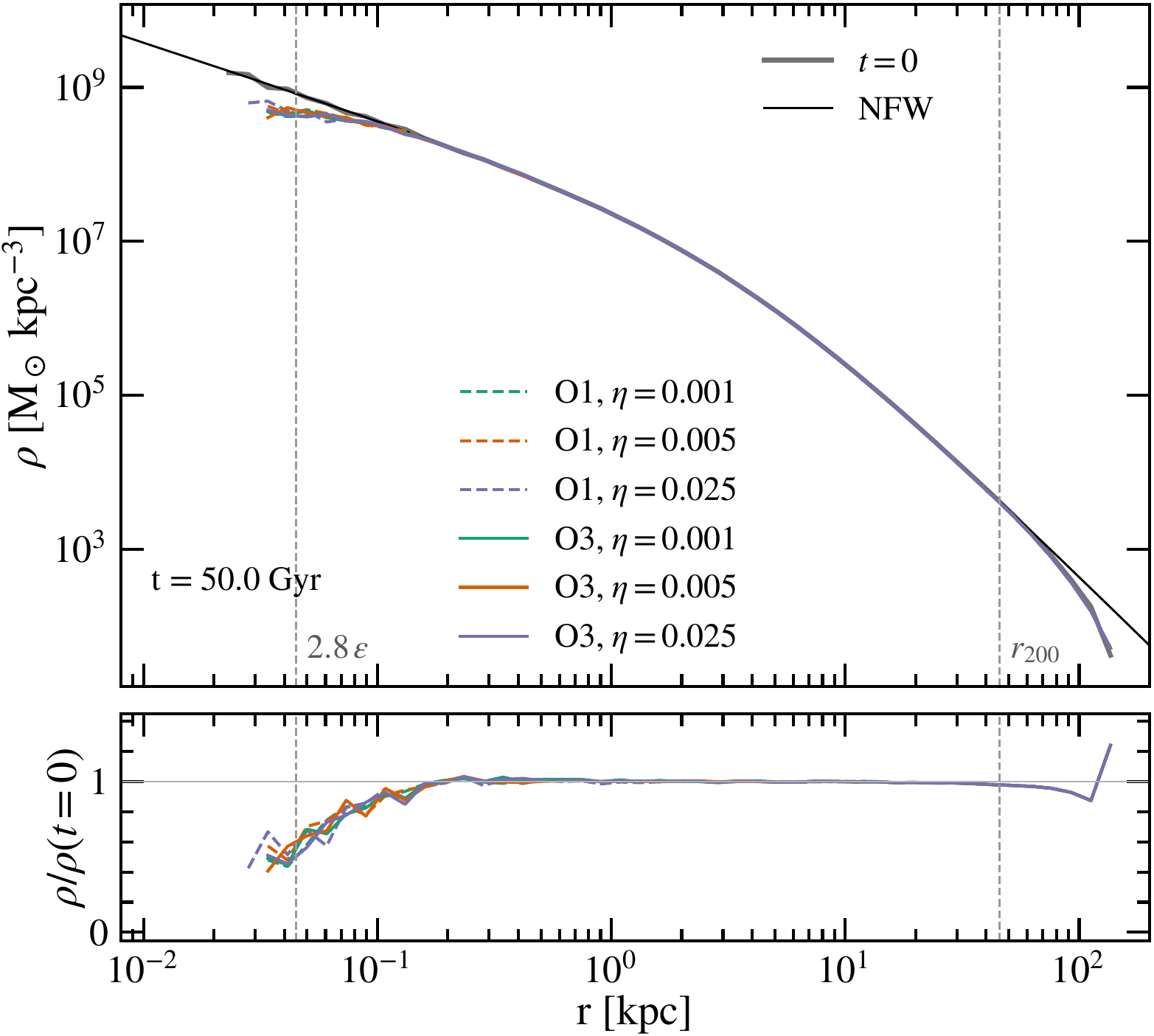}
    \hfill
    \includegraphics[width=0.49\textwidth]{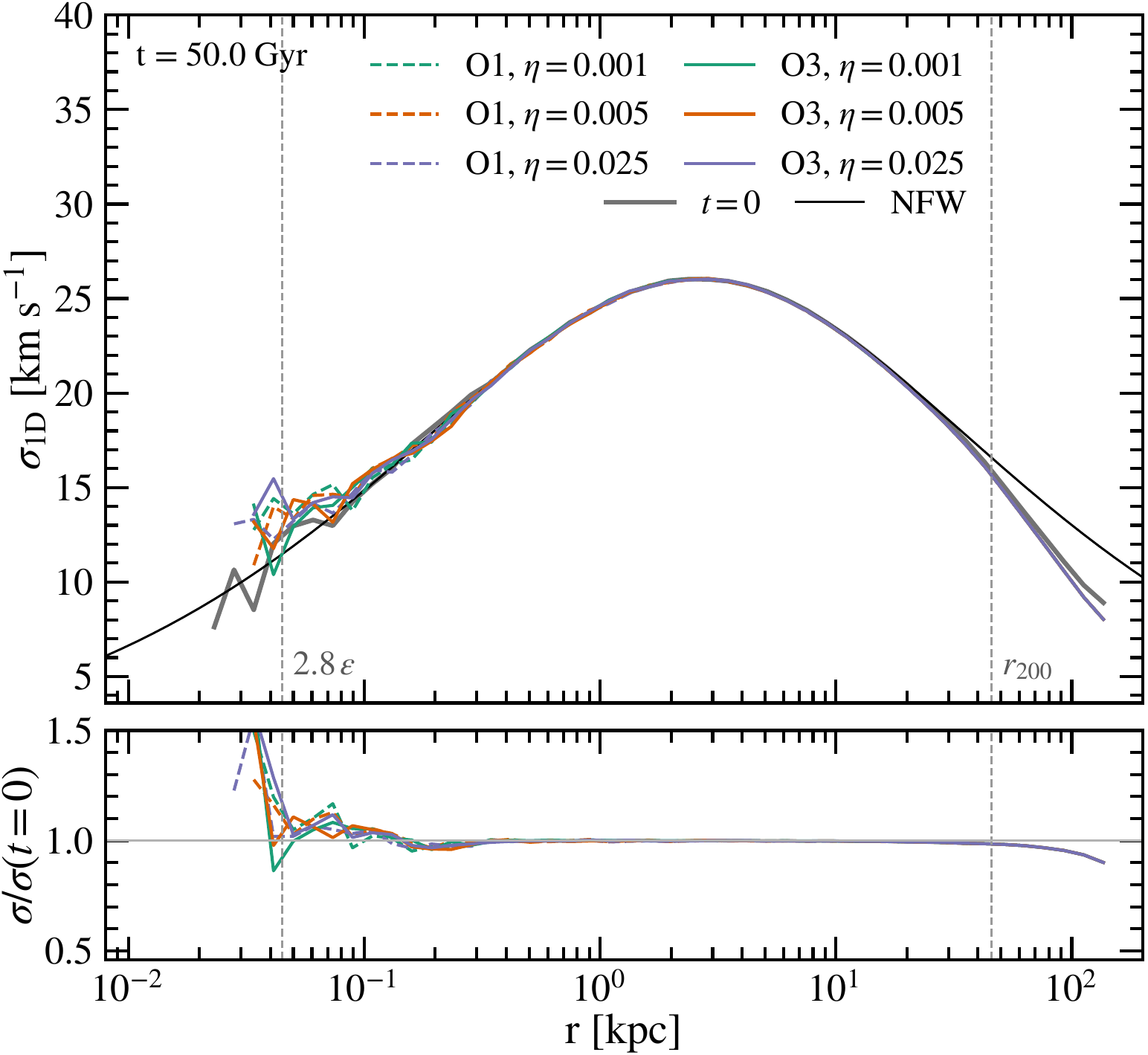}
    \caption{Collisionless evolution of the $M_{200}=10^{10}\,\mathrm{M_\odot}$
    NFW halo with $N_{200}=10^7$ particles and $\sigma/m=0$.
    Left: radial density profile after $50\,\mathrm{Gyr}$.
    Right: one-dimensional velocity dispersion at the same time.
    Dashed and solid curves show monopole (O1) and quadrupole (O3)
    gravity, respectively, for $\eta=0.001$ (green), $0.005$ (orange),
    and $0.025$ (purple). Thick grey curves show the initial conditions, and thin black curves the unsoftened, untruncated NFW density profile and its isotropic Jeans solution; the lower panels show ratios to the initial conditions. Vertical lines mark the spline-softening radius $2.8\epsilon$ and $r_{200}$. The deviations from the NFW curves beyond $r_{200}$ reflect the truncation of the initial conditions.}
    \label{fig:collisionless_1e7}
\end{figure*}

To quantify the effect of the time-stepping, we run the simulation with the
time-step parameter of \cref{eq:dtgrav} set to
$\eta = 0.001$, $0.005$, and $0.025$, each with O1 and with O3 gravity. We
use the fiducial opening parameters, $\alpha = 0.0045$ for O1 and
$\alpha = 0.0081$ for O3, so that both orders have the same force accuracy,
as shown in \cref{subsec:forceAccuracy}.
We use a maximum of $K=8$ particles per leaf. We measure density and velocity-dispersion profiles in logarithmically
spaced spherical shells between $0.5\,\epsilon$ and $150\,\mathrm{kpc}$
around the halo centre, which we locate with the shrinking-sphere method
\citep{Power2003}, and discard shells with fewer than 32 particles. The
one-dimensional velocity dispersion of a shell is
\begin{equation}
    \sigma_{\rm 1D} = \left[\frac{1}{3}\left\langle
    \left|\bm{v}-\langle\bm{v}\rangle\right|^{2}
    \right\rangle\right]^{1/2},
    \label{eq:sigma1d}
\end{equation}
where the averages run over the particles of the shell.
\Cref{fig:collisionless_1e7} shows the radial density and
velocity-dispersion profiles after $50\,\mathrm{Gyr}$. All six runs keep the
NFW profile over most of the halo. Inside $r \approx 200\,\mathrm{pc}$, the density falls increasingly below its initial profile, to about half of it at the innermost resolved radii, and the dispersion is correspondingly enhanced, as expected from two-body relaxation \citep{Power2003}.

\begin{figure}
    \centering
    \includegraphics[width=1\linewidth]{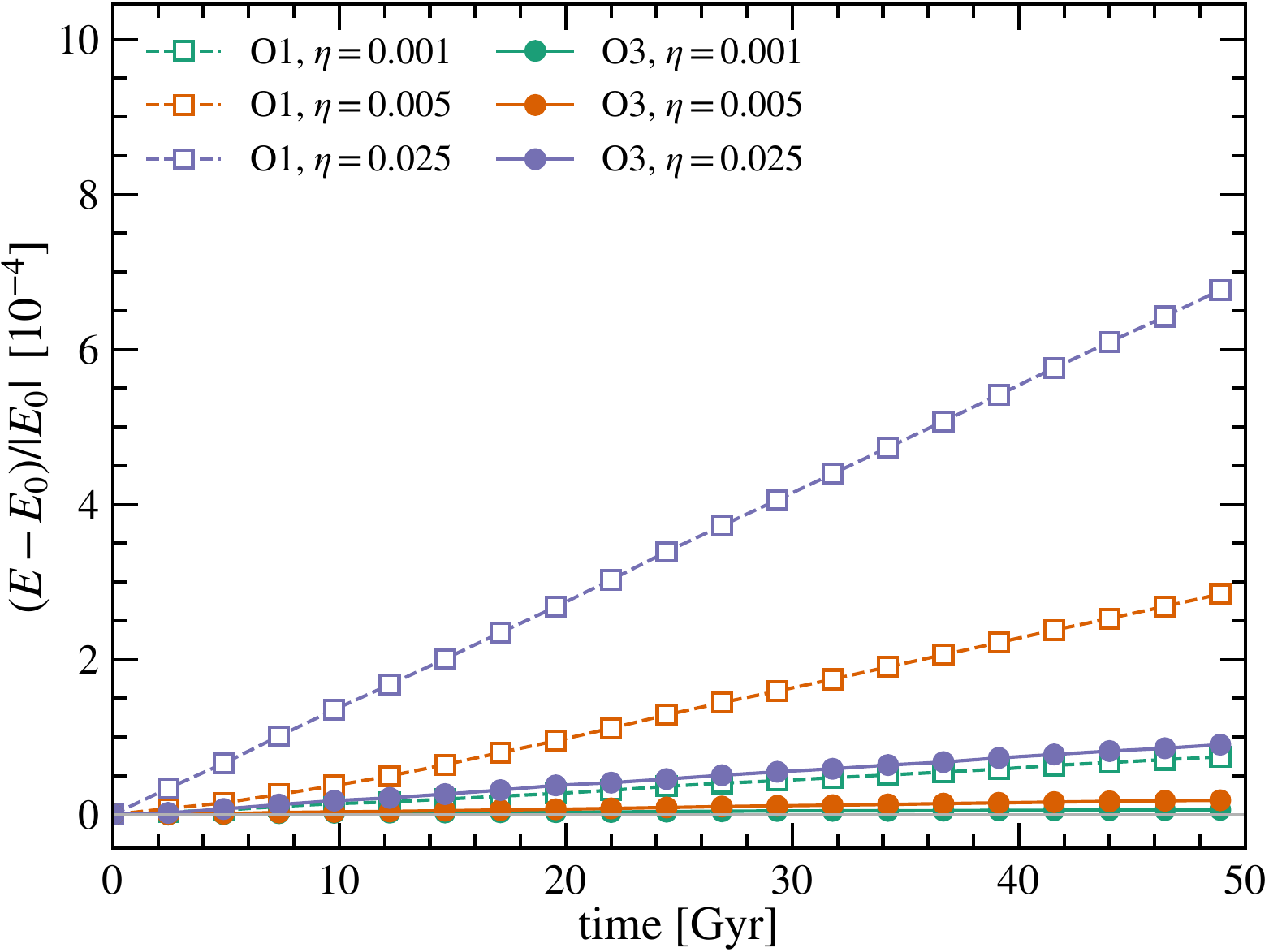}
    \caption{Relative change in the total energy,
    $(E-E_0)/|E_0|$ in units of $10^{-4}$, during the $50\,\mathrm{Gyr}$
    collisionless evolution shown in \cref{fig:collisionless_1e7}, where
    $E_0$ is the initial kinetic plus potential energy. The potential
    energy is computed by direct summation for the initial conditions and
    20 snapshots (symbols), the last at $49\,\mathrm{Gyr}$. Green, orange, and purple curves show
    $\eta=0.001$, $0.005$, and $0.025$, respectively, and dashed and solid
    curves denote O1 and O3 gravity, respectively.}
    \label{fig:collisionless_1e7_energy}
\end{figure}

\Cref{fig:collisionless_1e7_energy} shows the evolution of the total
(kinetic plus potential) energy. We compute the potential energy by direct
summation over all particle pairs for the initial conditions and for 20
snapshots, because the on-the-fly estimate of the tree is too noisy to show
the secular evolution, as we discuss in \cref{app:ImpactEnergyMeasurement}.
In the collisionless runs, energy conservation improves with decreasing $\eta$. At the same force accuracy, the O3 runs drift by less than $10^{-4}$ after $49\,\mathrm{Gyr}$, the last epoch with a direct sum, roughly an order of magnitude less than the O1 runs at the same $\eta$, which drift by $(0.8$--$7)\times10^{-4}$.

\subsection{Isolated core collapse}
\label{subsec:coreCollapse}
To test self-interactions and gravity together, we simulate the gravothermal collapse of the halo in \cref{subsec:collisionlessEquilibrium} with
$\sigma/m = 30\,\mathrm{cm^{2}\,g^{-1}}$. We repeat the six runs of that
section, O1 and O3 gravity with $\eta = 0.001$, $0.005$, and $0.025$. All SIDM runs
of this paper use $C_{\rm SIDM}=0.1$ (\cref{eq:sidm_timestep}). This
is the test we performed in \citet{ZierSidm2026}, and we take the
\arepoTwo run of that paper as the comparison run. That run uses $\eta=0.005$, $\alpha=0.0025$ with one particle per leaf, $32\pm5$ neighbours, the same softening, double precision, and the SIDM time-step criterion of \citet{ZierSidm2026}.

We measure the central density $\rho_{\rm c}$ with the estimator of \citet{ZierSidm2026},
adapted from \citet{Tran2025_2}, but with smaller particle thresholds. After locating the halo centre with a shrinking-sphere
iteration \citep{Power2003}, we grow a central sphere shell by shell, each shell containing at
least 100 particles. Once the sphere contains 900 particles, we stop adding
shells as soon as the density of the next shell deviates from the mean
density of the sphere by more than twice their combined Poisson
uncertainty, and $\rho_{\rm c}$ is the mean density of the sphere at that
point. We apply the same estimator with the same settings to the \arepoTwo run.

\begin{figure}
    \centering
    \includegraphics[width=1\linewidth]{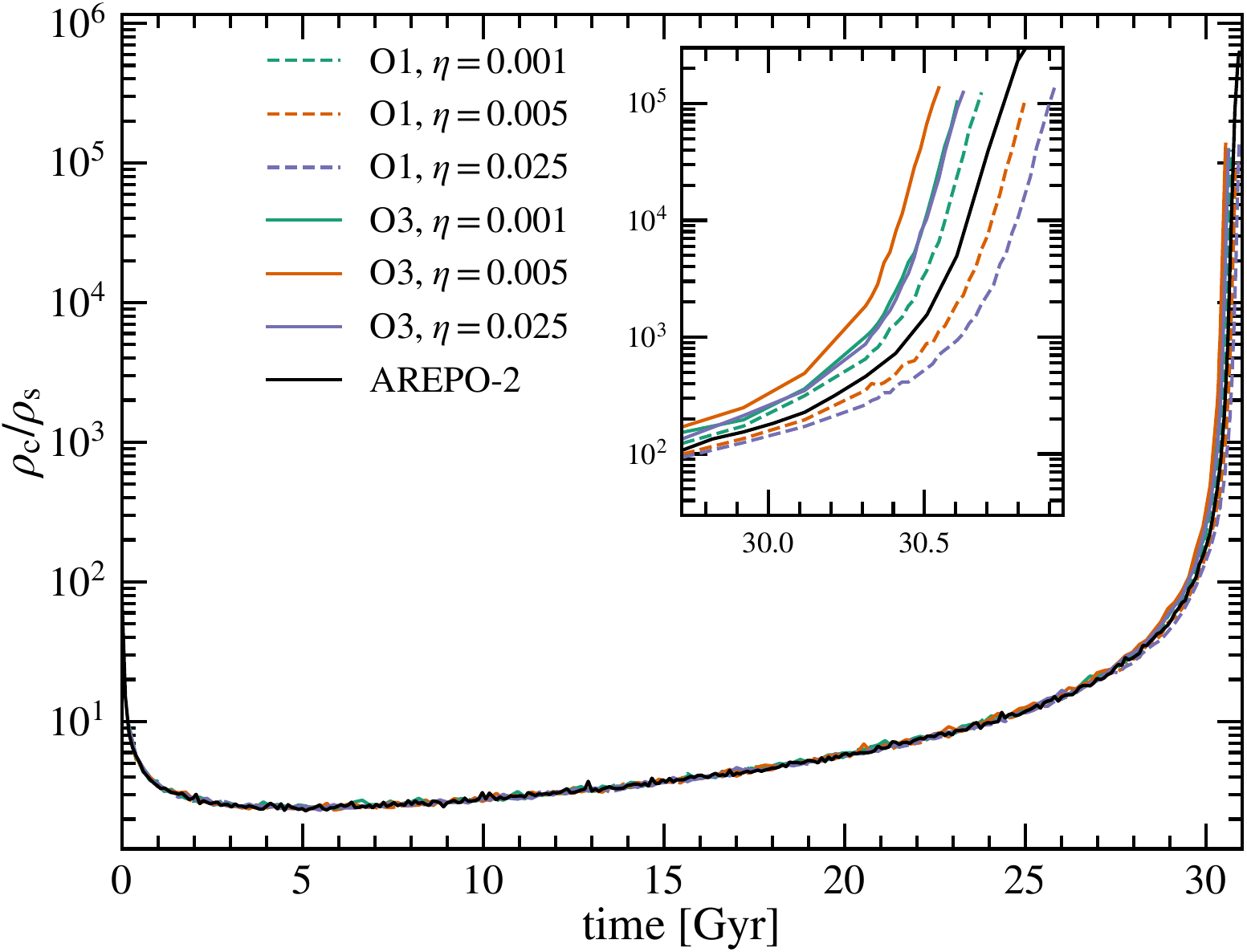}
    \caption{Evolution of the central density, in units of the NFW scale
    density $\rho_{\rm s}$, during gravothermal collapse of the
    $M_{200}=10^{10}\,\mathrm{M_\odot}$ halo with $N_{200}=10^7$ particles and
    $\sigma/m=30\,\mathrm{cm^2\,g^{-1}}$. The \nori runs use
    $\eta=0.001$ (green), $0.005$ (orange), and $0.025$ (purple),
    with O1 gravity shown by dashed curves and O3 by solid curves.
    The black curve is the \arepoTwo comparison run from
    \citet{ZierSidm2026}. The inset enlarges the final collapse.}
    \label{fig:collapse_1e7_core}
\end{figure}

\Cref{fig:collapse_1e7_core} shows the evolution of the central density.
All runs agree closely throughout the evolution, and the times at which they
reach a given central density during the final collapse differ by a small
fraction of the collapse time. The O3 runs collapse slightly earlier than the
O1 runs and the \arepoTwo run. Among the O1 runs, a larger $\eta$ delays
the collapse.

\begin{figure}
    \centering
    \includegraphics[width=1\linewidth]{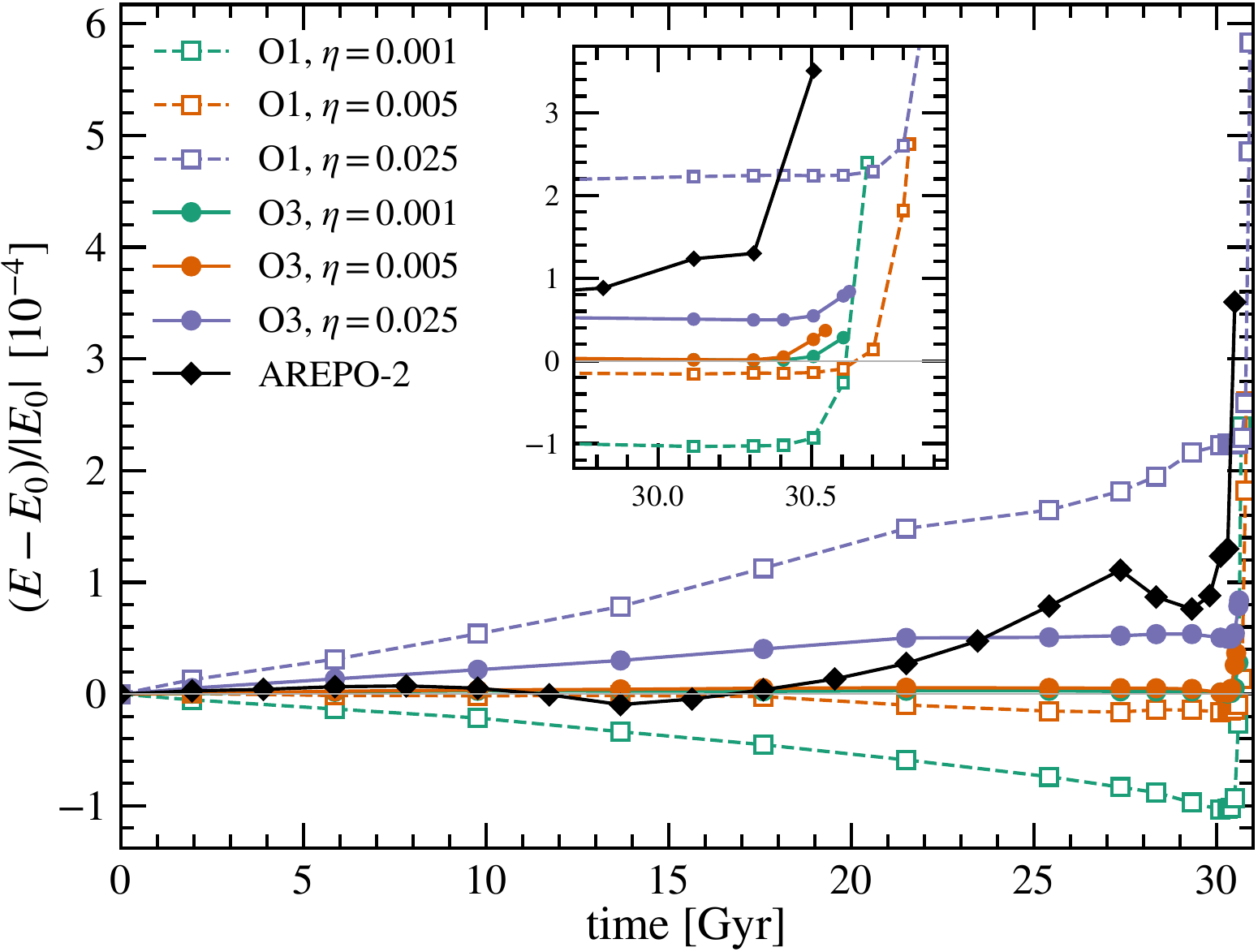}
    \caption{Relative change in the total energy,
    $(E-E_0)/|E_0|$ in units of $10^{-4}$, for the core-collapse
    simulations in \cref{fig:collapse_1e7_core}, with the same colours
    and line styles. Symbols mark the snapshots with a direct-summation
    potential energy; $E_0$ is the energy of the initial conditions. The last
    direct-summation measurement for \arepoTwo is at $30.5\,\mathrm{Gyr}$, before its
    central density reaches $10^{4}\,\rho_{\rm s}$.}
    \label{fig:collapse_1e7_energy}
\end{figure}

In the late stages of the gravothermal collapse, energy conservation becomes an important
diagnostic. \Cref{fig:collapse_1e7_energy} shows the evolution of the relative
energy error, with the potential energy from direct summation as in \cref{subsec:collisionlessEquilibrium}.
Up to the final collapse, the O3 runs conserve energy considerably better than
the O1 runs at the same force accuracy, as in the collisionless test, whereas the
drift of the O1 runs is not monotonic in $\eta$. The \arepoTwo run with individual
time-steps conserves energy well for the first $20\,\mathrm{Gyr}$, after which its
error grows up to its last direct-summation epoch, shortly before the final collapse.
In the final collapse, the energy of the O1 runs rises further, whereas the O3 runs stay within $10^{-4}$ of their initial energy up to the highest central densities reached.

\section{Code performance}
\label{sec:performance}

We measure the performance of \nori using the isolated NFW halo introduced in
\cref{subsec:collisionlessEquilibrium}.
We vary the particle number within $r_{200}$ from $10^5$ to $10^9$ and scale the softening approximately with $N^{-1/3}$, which
gives values between $0.074\,\mathrm{kpc}$ and $0.004\,\mathrm{kpc}$.
All tests are performed on the Cannon cluster at Harvard University. The CPU
compute nodes contain two 56-core Intel Sapphire Rapids processors and therefore 112
cores in total. Each GPU compute node combines 64 Intel Ice Lake CPU cores with four
NVIDIA A100 GPUs with $80\,\mathrm{GB}$ of device memory. We use one MPI rank
per GPU.

Unless mentioned otherwise, the \nori simulations use O3 gravity with
$\alpha=0.0081$, $\eta=0.005$, $K=8$, $C_{\rm SIDM}=0.1$, and
$\sigma/m=30\,\mathrm{cm^2\,g^{-1}}$. The \arepoTwo comparison uses the standard monopole implementation of
gravity, $\alpha=0.0025$, one particle per leaf, and a forced global
time-step. As described in \cref{subsec:forceAccuracy}, \nori with O3, $K=8$, and $\alpha=0.0081$ matches the
90th-percentile force error of \nori with O1, $K=1$, and $\alpha=0.0025$ in the Hernquist test. The latter
uses the gravity settings of the \arepoTwo comparison. O1 with $K=8$ reaches the same error at $\alpha=0.0045$, the value used in \cref{subsec:impactBucketSize}.

Each benchmark evolves the initial halo for 64 global time-steps of equal length. We
discard the first ten steps and calculate the median time over steps 10 to 63.

\subsection{Strong scaling}
\label{subsec:strongScaling}
\begin{figure*}
    \centering
    \includegraphics[width=0.48\linewidth]{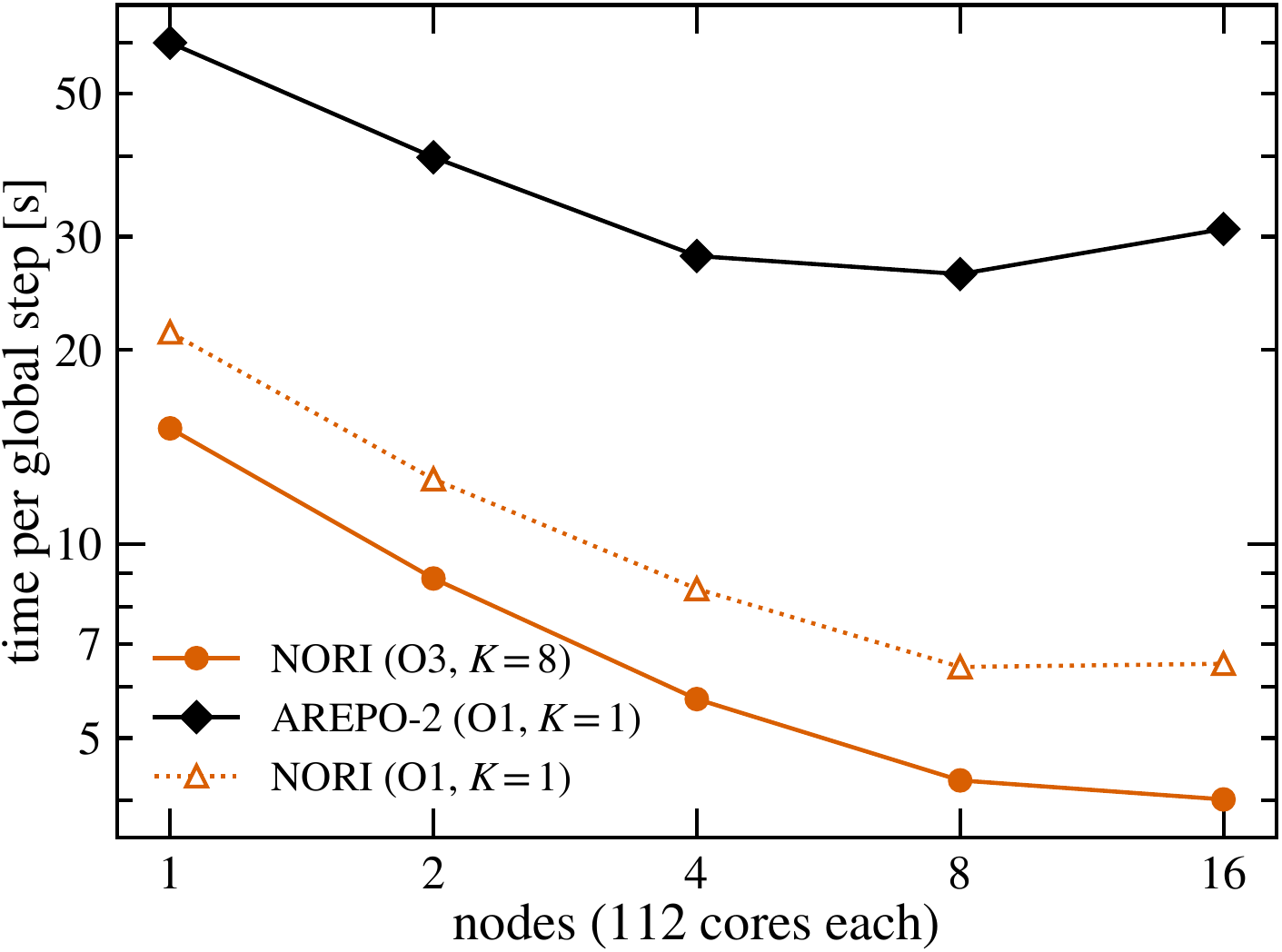}
    \includegraphics[width=0.48\linewidth]{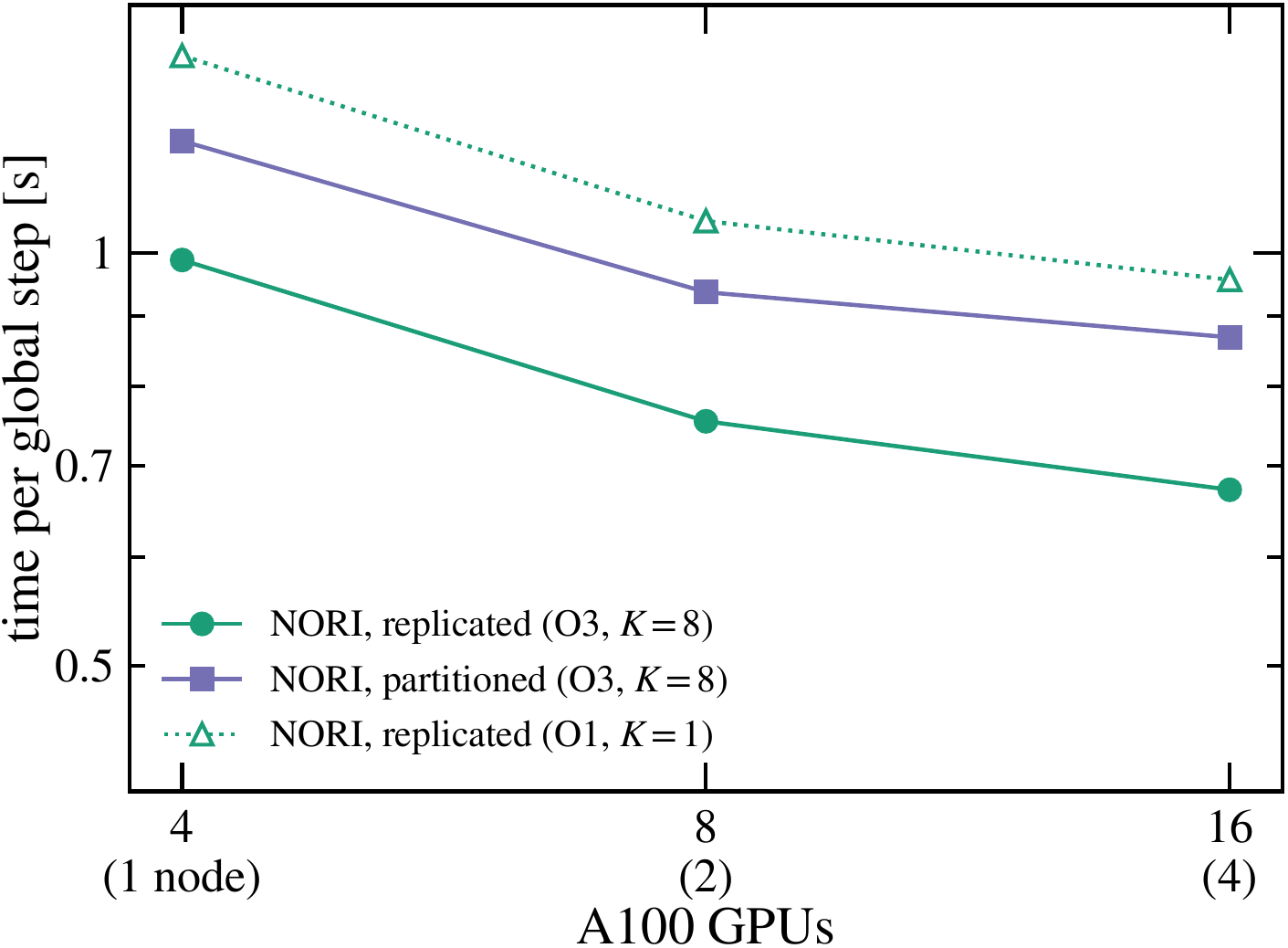}
    \caption{Strong scaling for the isolated SIDM halo with
    $N_{200}=10^8$ particles. Symbols give the median wall-clock
    time per global time-step.
    Left: \nori (orange) and \arepoTwo (black) on 1--16 CPU nodes,
    each with 112 cores. Right: \nori
    on 1--4 GPU nodes, each containing four A100 GPUs, in
    the replicated GPU mode (green) or the partitioned GPU mode
    (purple). Dotted lines with open triangles show \nori in the \arepoTwo configuration (O1, $K=1$, $\alpha=0.0025$). Both codes use global time-steps in this
    test.}
    \label{fig:strongScaling}
\end{figure*}
We first perform a strong-scaling test with $10^{8}$ particles within $r_{200}$, varying
the number of compute nodes. The results for both CPU and GPU runs are shown in
\cref{fig:strongScaling}. 

On one CPU node, a step takes $15\,\mathrm{s}$ with \nori and $60\,\mathrm{s}$
with \arepoTwo, a speed-up of around 4. The speed-up grows with the number of nodes
and reaches 7.7 on 16 nodes: from one to 16 nodes, the time per step of \nori
falls by a factor of 3.8, whereas \arepoTwo gains only a factor of 2.3 up to
eight nodes and becomes slower again on 16, where communication, waiting for the slowest rank, and the domain decomposition outweigh the shorter tree walks. With its dynamic load balancing
within each CPU node, \nori still improves slightly from eight to 16 nodes.
Part of the speed-up comes from the gravity settings: with the \arepoTwo
settings (O1, $K=1$, $\alpha=0.0025$), \nori is 2.8 and 4.7 times faster than
\arepoTwo on one and 16 nodes, and the fiducial settings (O3, $K=8$, $\alpha=0.0081$) contribute a further
factor of 1.4 on one node, 1.6 on 16 nodes, and 1.4 on GPUs. This factor comes
from the larger leaves and the larger $\alpha$ they permit at the same force
error rather than from the quadrupole: O1 with $K=8$ and $\alpha=0.0045$ is slightly
faster than O3 (\cref{subsec:impactBucketSize}).

On GPUs, one node, where a step takes $0.99\,\mathrm{s}$ in the replicated mode,
is 15 times faster than one CPU node running \nori and 60 times faster than one
CPU node running \arepoTwo. Four GPU nodes are 1.5 times faster than one in the
replicated mode and 1.4 times faster in the partitioned mode, so that four GPU
nodes are 6 times faster than 16 CPU nodes running \nori. We do not expect
ideal strong scaling, because every GPU repeats the sorting, tree construction,
and other linear operations, the accelerations have to be communicated after
the walk, and the targets are divided between the GPUs by a cost-weighted partition
that is fixed between reorderings, without the dynamic chunk scheduling of the
CPU ranks (\cref{subsec:parallelization}). The partitioned mode
is about a quarter slower than the replicated mode at every node count. This
additional cost comes from the slower access to the particles held by the
other GPUs in all kernels, most of all in the tree construction and the
neighbour search, and from the synchronization across the node.

\subsection{Scaling with problem size}
\label{subsec:scalingProblemSize}
\begin{figure*}
    \centering
    \includegraphics[width=0.48\linewidth]{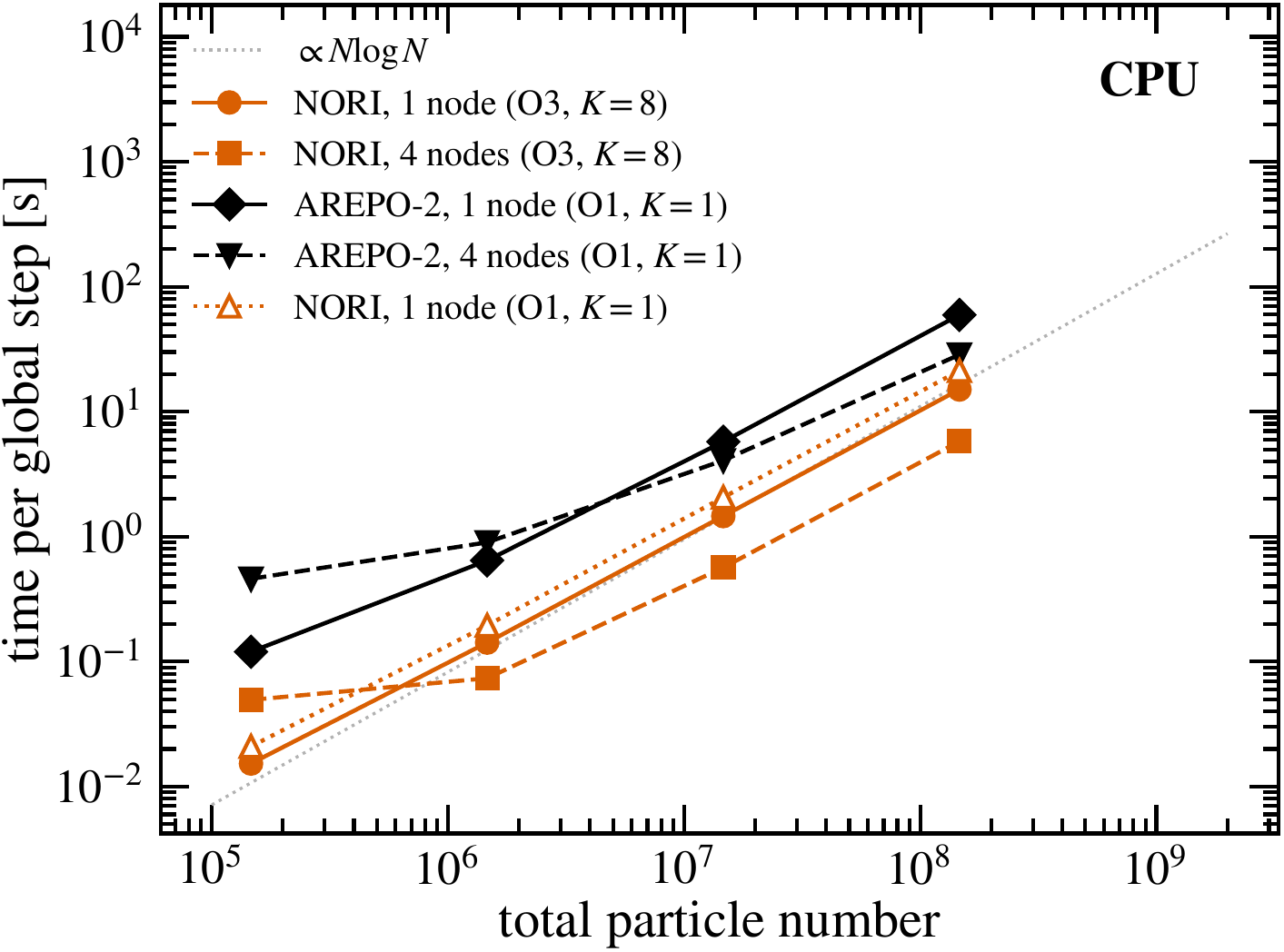}
    \includegraphics[width=0.48\linewidth]{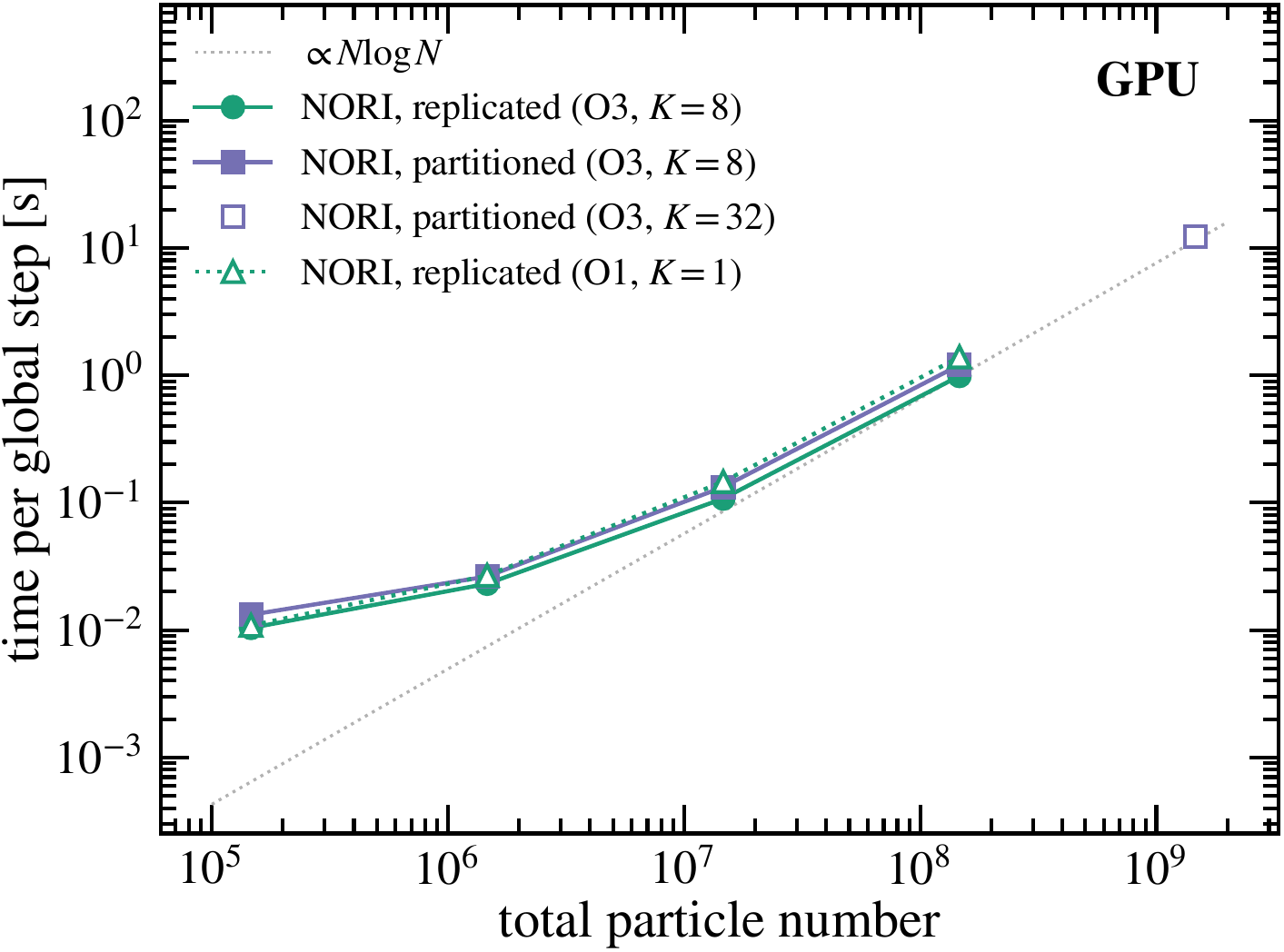}
    \caption{Dependence of the median wall-clock time per global time-step
    on the total particle number $N$, including particles outside
    $r_{200}$. Left: \nori (orange) and \arepoTwo (black) on
    one node (solid) and four CPU nodes (dashed).
    Right: the replicated (green) and partitioned (purple) GPU
    modes on one node with four A100 GPUs. Dotted lines with open triangles show \nori in the \arepoTwo configuration (O1, $K=1$, $\alpha=0.0025$). Grey dotted lines
    indicate $N\log N$ scaling as a guide to the eye.
    The partitioned mode permits the largest run,
    $N\simeq1.47\times10^9$, for which a complete replica
    does not fit on a single GPU. This run (open square) uses $K=32$, since
    the tree for $K=8$ would exceed the $80\,\mathrm{GB}$ of device memory;
    at $N_{200}=10^8$ in the replicated mode, $K=32$ is about 15 per cent slower than $K=8$
    (\cref{subsec:impactBucketSize}).}

    \label{fig:problem_size}
\end{figure*}
Next, we keep the number of nodes constant and vary the problem size from $10^5$ to $10^9$ particles within $r_{200}$, as shown in \cref{fig:problem_size}.
For small particle numbers, kernel launch overhead,
MPI collectives, and tree construction dominate the run time. One GPU node in the replicated mode takes
$0.01\,\mathrm{s}$ per step at $N_{200}=10^5$ and
$0.02\,\mathrm{s}$ at $N_{200}=10^6$. At larger $N$, the measurements
approach the expected $N\log N$ scaling of the tree algorithm. On four CPU
nodes and for $N_{200}=10^8$, \nori is 4.9 times faster than \arepoTwo.

\subsection{Impact of the leaf size}
\label{subsec:impactBucketSize}
\begin{figure}
    \centering
    \includegraphics[width=1\linewidth]{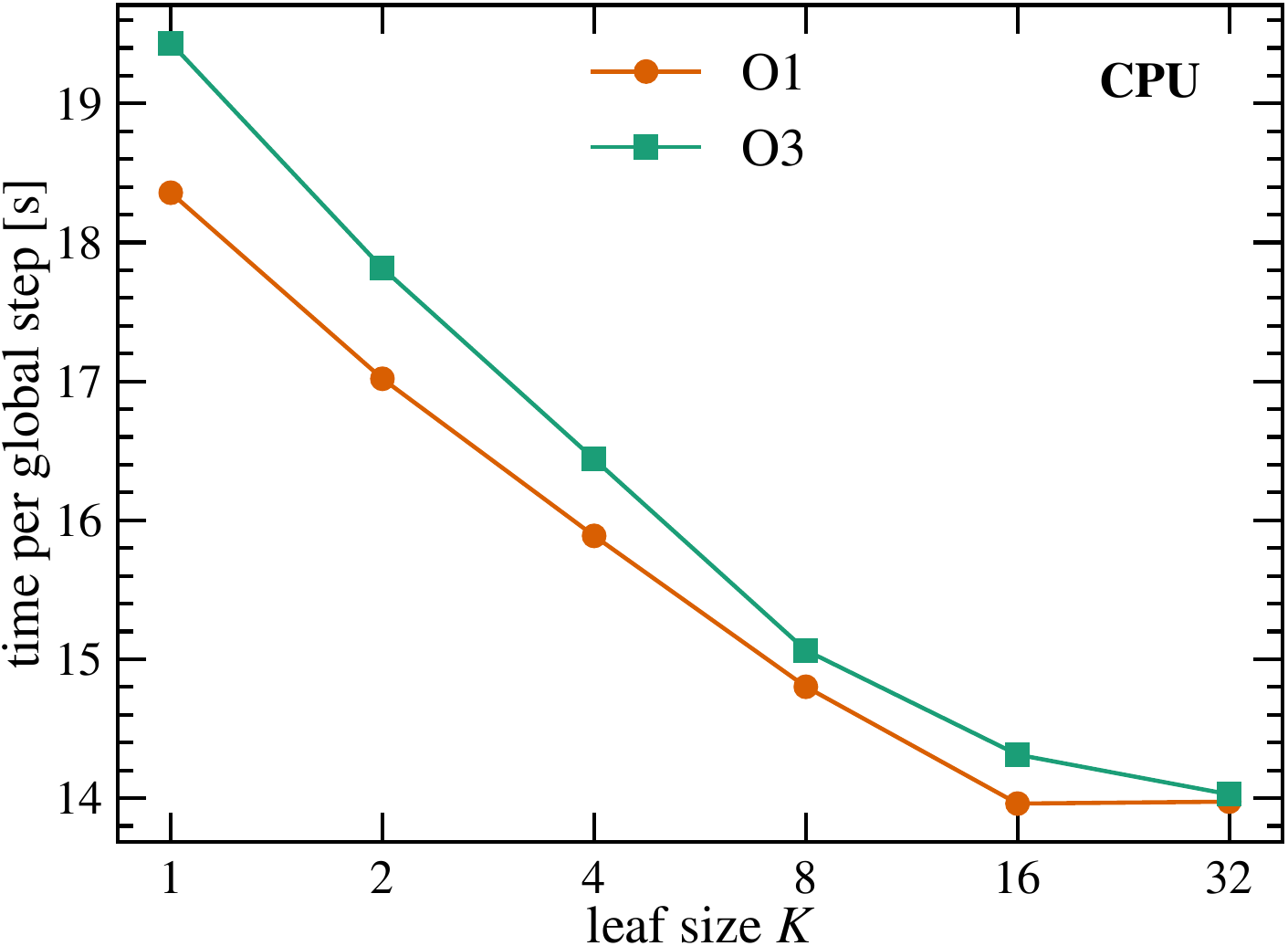}
    \par\vspace{3mm} 
    \includegraphics[width=1\linewidth]{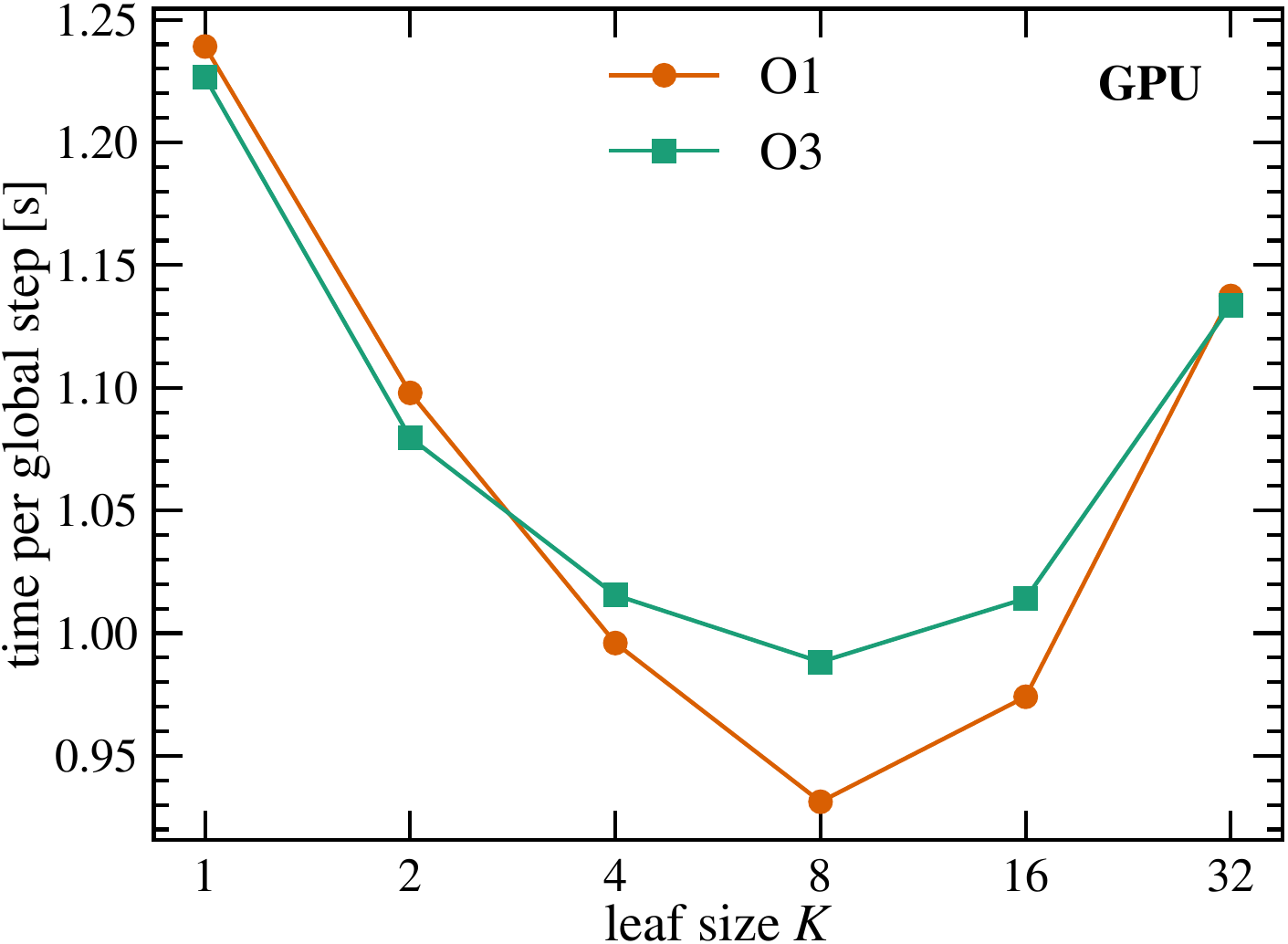}
    \caption{Wall-clock cost of a global time-step as a function of the
    maximum number of particles per leaf, $K$, for the
    $N_{200}=10^8$ SIDM halo. Top: one CPU node with 112 cores.
    Bottom: one GPU node with four A100 GPUs. Symbols show
    median timings for monopole gravity with $\alpha=0.0045$ (O1, orange)
    and quadrupole gravity with $\alpha=0.0081$ (O3, green), the opening
    parameters at which the two orders have the same force error at $K=8$
    (\cref{subsec:forceAccuracy}), held fixed as $K$ varies. On the CPU node,
    the cost falls with $K$ and flattens at $K=16$--32; on the GPU node, both
    orders are fastest at $K=8$.}
    \label{fig:bucket_size}
\end{figure}

The maximum number of particles per leaf, $K$, is a free parameter of the
tree. Larger leaves reduce the number of nodes, and with it the construction
cost and the memory use, but more pair interactions are then evaluated
directly instead of through a multipole expansion
(\cref{subsec:treeImplementation}). \Cref{fig:bucket_size} shows the median
cost of a global time-step as a function of $K$ for the $N_{200}=10^8$ halo
on one CPU node and on one GPU node, for O1 with $\alpha=0.0045$ and O3 with
$\alpha=0.0081$, so that the two orders have the same 90th-percentile force
error at $K=8$ (\cref{subsec:forceAccuracy}). Within each order, $\alpha$ is
held fixed, so that the error falls with $K$, by a third to a half from $K=1$
to $K=32$ (\cref{fig:forceError}).

On CPUs, the cost falls with $K$ and flattens at $K=16$--32. On GPUs, both
orders are fastest at $K=8$, and the cost rises again for larger leaves,
because the direct particle interactions in the leaf loops keep only a small
fraction of the threads of a warp busy (\cref{subsec:parallelization}). At
this force error, O1 is a few per cent cheaper than O3 on the CPU node at
every $K$ and on the GPU node around the optimum: O3 opens fewer nodes
(\cref{subsec:forceAccuracy}), but each accepted node costs more
(\cref{subsec:gravityImplementation}). The gain of the fiducial configuration
over the \arepoTwo settings in \cref{subsec:strongScaling} therefore comes from
$K=8$ and the larger $\alpha$ it permits at the same error, not from the
quadrupole. We keep O3 as the fiducial order for its better energy
conservation (\cref{sec:validation}). Larger values of $K$ nevertheless reduce
the memory of the tree and can be preferable for the largest simulations on
GPUs; the run with $1.47\times10^9$ particles in
\cref{subsec:scalingProblemSize} uses $K=32$.

\subsection{Time to solution for a full collapse}
\label{subsec:performanceFullSimulation}

The previous benchmarks used global time-steps for both codes, but
production runs of \arepoTwo use individual time-steps. To assess performance
in production, we compare the run time of the fiducial core-collapse
simulation of \cref{subsec:coreCollapse} (O3, $\eta=0.005$, $10^7$ particles
within $r_{200}$) with \nori on one four-A100 GPU node and on one CPU node against
\arepoTwo with its individual time-steps on one CPU node.

\begin{figure}
    \centering
    \includegraphics[width=\linewidth]{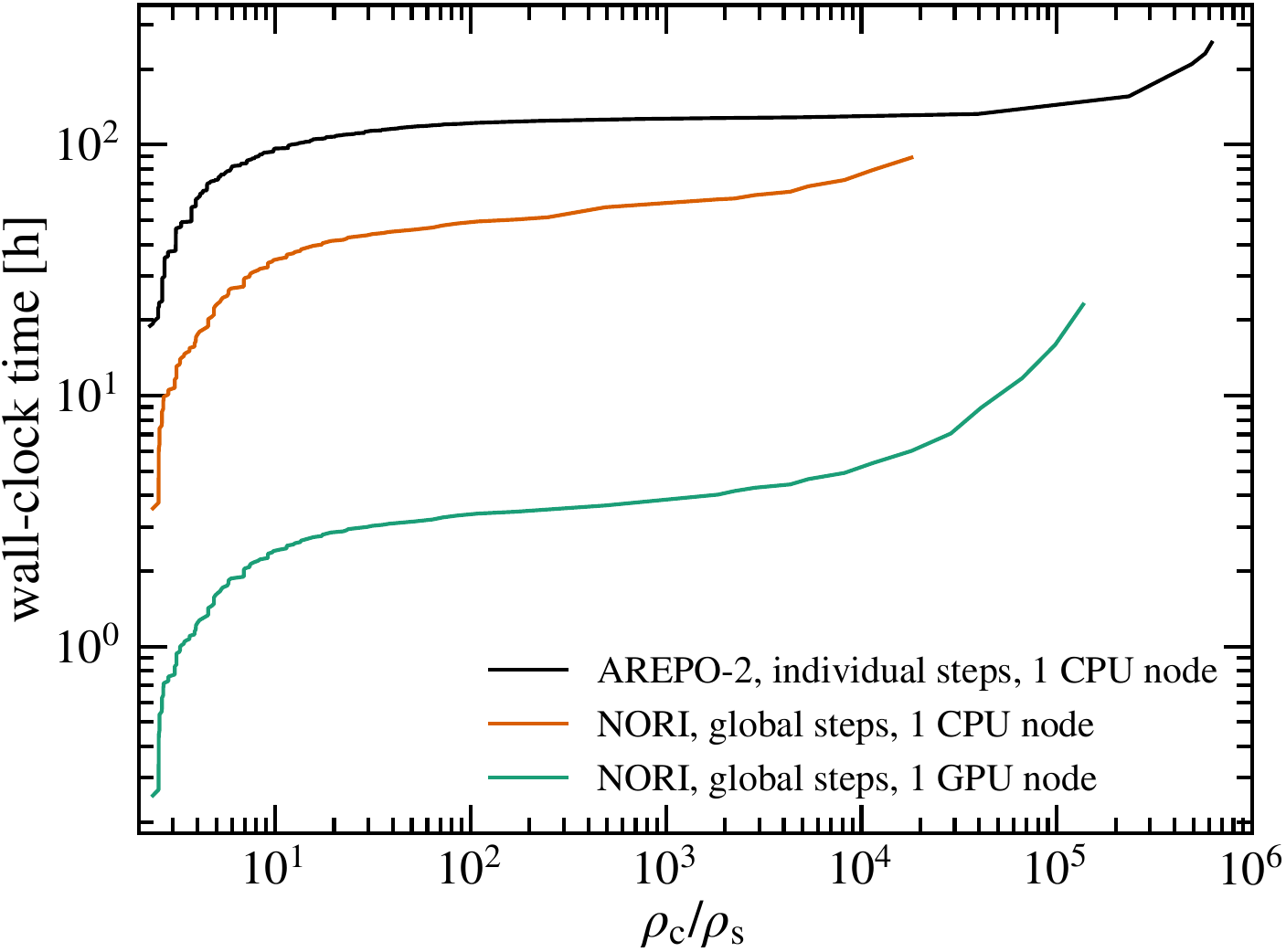}
    \caption{Cumulative wall-clock time to reach a given central density
    during collapse of the $N_{200}=10^7$ SIDM halo.
    We compare \arepoTwo with individual time-steps on one
    CPU node (black) to \nori with global time-steps on one
    CPU node (orange) and one four-A100 GPU node (green).
   The GPU run reaches a given density
    at substantially lower wall-clock cost, although the
    advantage decreases during runaway collapse.}
    \label{fig:wallclock_vs_density}
\end{figure}

\Cref{fig:wallclock_vs_density} shows the wall-clock time at which each run reaches a given central density.
On the GPU node, \nori reaches $\rho_{\rm c}/\rho_{\rm s}=10^{3}$ and $10^{4}$ after $3.8$ and $5.2\,\mathrm{h}$,
whereas \arepoTwo needs $5.3$ and $5.4$ days on the CPU node. The advantage of the GPU node
shrinks from a factor of $34$ at $10^{3}$ to $25$ at $10^{4}$, and it keeps shrinking in the runaway
collapse, to $18$ at $3\times10^{4}$ and $9$ at $10^{5}\,\rho_{\rm s}$: the global time-step advances every particle at the rate required by the core, including the particles outside $r_{200}$, while the individual time-steps of \arepoTwo allow the particles of the outer halo to advance less frequently. On the CPU node,
\nori reaches $\rho_{\rm c}/\rho_{\rm s}=10^{2}$, $10^{3}$, and $10^{4}$ after $49$, $58$, and $76\,\mathrm{h}$,
which is $2.5$, $2.2$, and $1.7$ times faster than \arepoTwo. Between $10^{3}$ and $10^{4}\,\rho_{\rm s}$, however,
\nori on the CPU node advances about 7 times more slowly than \arepoTwo, so that \arepoTwo overtakes it in the runaway
collapse; the \nori run on the CPU node ends at $\rho_{\rm c}/\rho_{\rm s}\approx2\times10^{4}$. On the GPU node, \nori is
still $1.7$ times faster than \arepoTwo over this interval, and \arepoTwo advances faster only above about
$3\times10^{4}\,\rho_{\rm s}$.

\section{A reference solution for SIDM gravothermal collapse}
\label{sec:referenceSolution}
 The efficiency of \nori allows us to simulate the gravothermal collapse with $N_{200}=10^8$ particles ($\approx1.5\times10^8$ in total). This is well below the limit of \nori but about three times as many as in the largest previous isolated collapse simulation \citep{Fischer2025Collapse} and ten times as many as in the
collapse runs of \cref{subsec:coreCollapse}.
 This calculation provides a higher-resolution comparison for the central-density evolution and a reference solution against which other codes can be tested. Because the particle number, softening, force tolerance, and time-step parameter all change relative to the fiducial calculation, the comparison does not isolate convergence with particle number.

For this reference solution, we repeat the core-collapse simulation of \cref{subsec:coreCollapse} with $N_{200}=10^8$ particles. The halo and the cross-section are the same as before. The softening is reduced to $\epsilon=0.00743\,\mathrm{kpc}$,
which keeps it proportional to the mean particle separation, and we use
quadrupole gravity with $\alpha=4\times10^{-4}$. The time-step parameter is $\eta=0.001$, and the leaf size is
$K=8$. We run the simulation twice from the same initial conditions, with two different seeds of the random-number generator, which we call seeds A and B. The seed enters the key of the random-number generator, so the two runs use independent random numbers for the scatterings. The difference illustrates the sensitivity of SIDM halo evolution to the randomness of Monte-Carlo scattering events.

\begin{figure}
    \centering
    \includegraphics[width=1\linewidth]{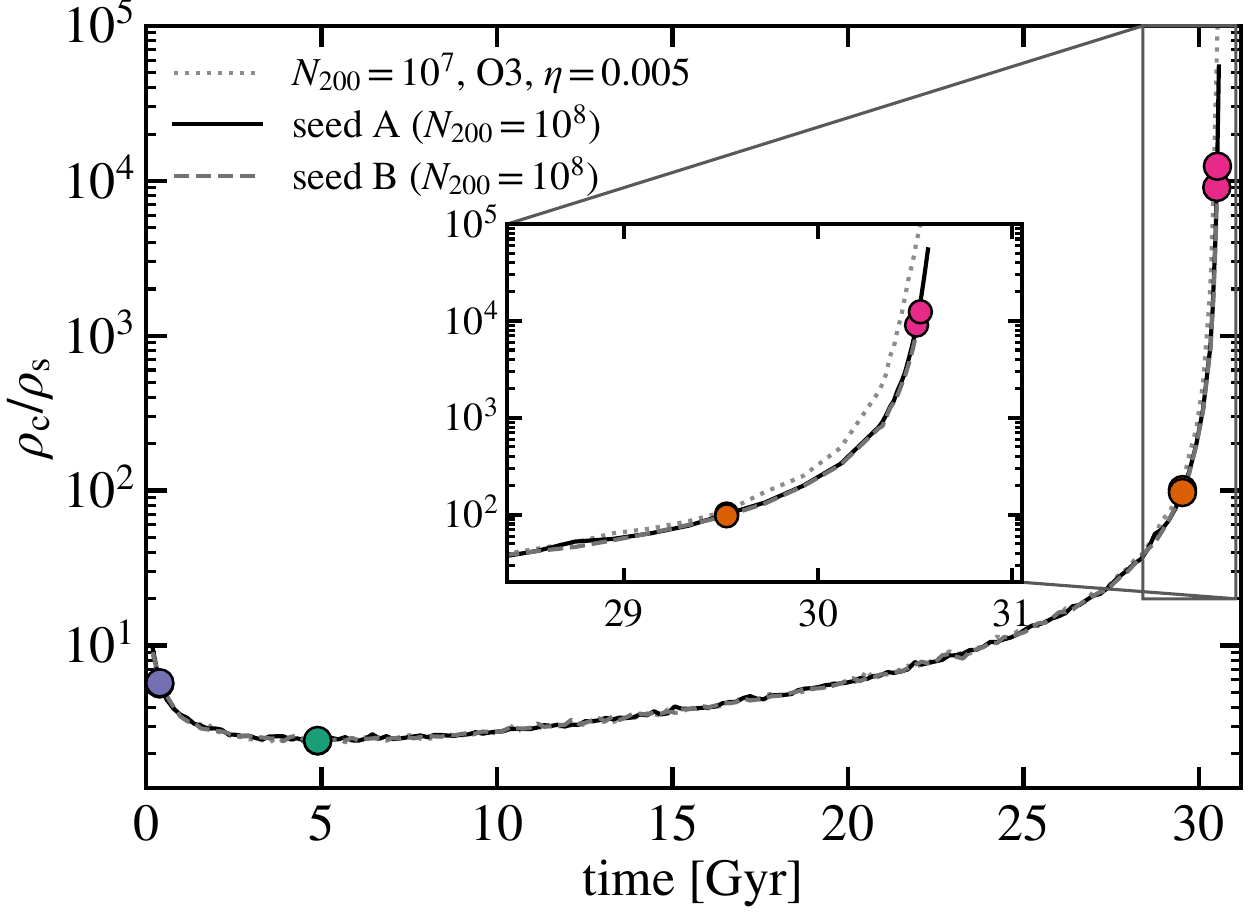}
      \includegraphics[width=1\linewidth]{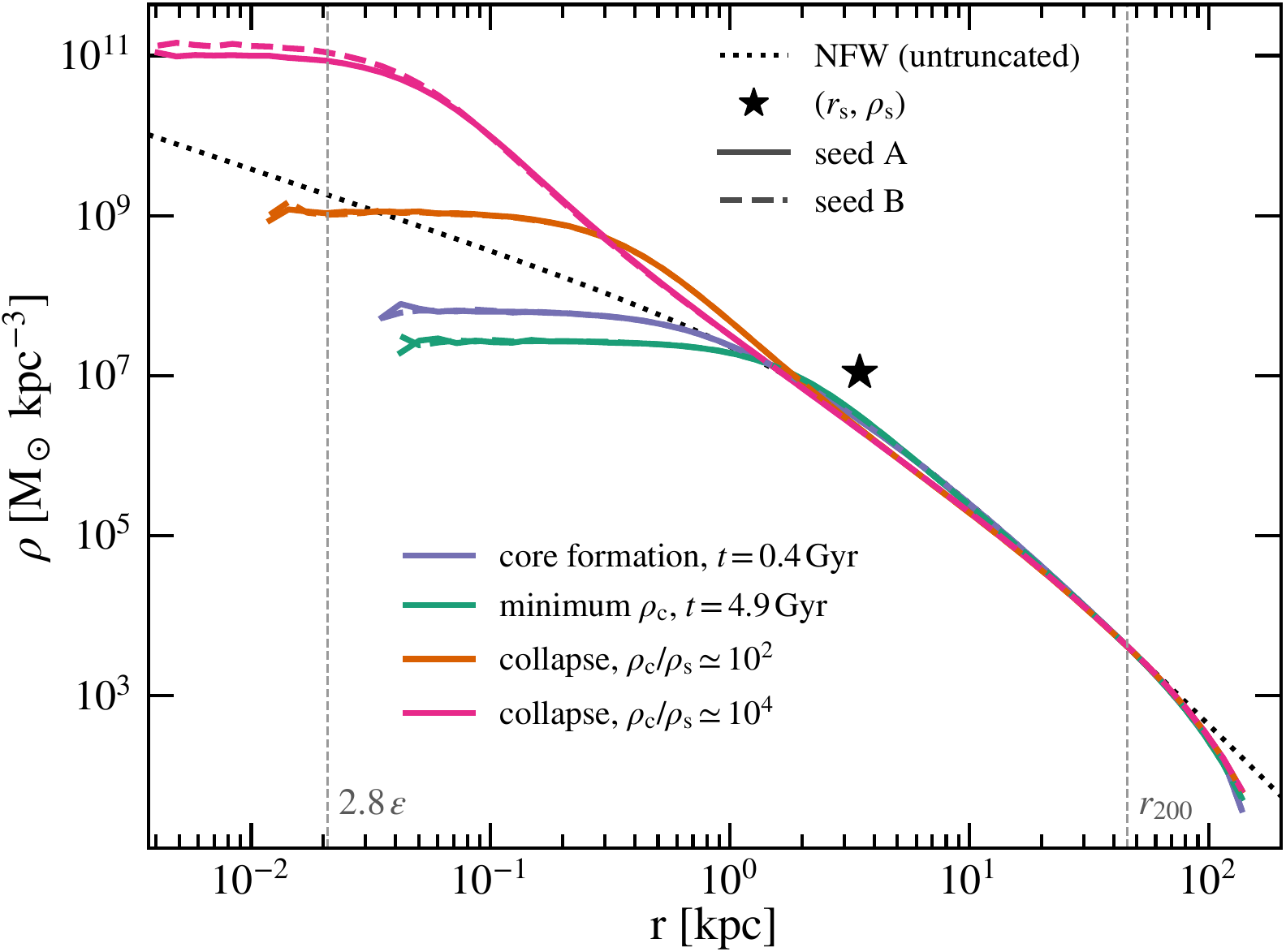}
        \includegraphics[width=1\linewidth]{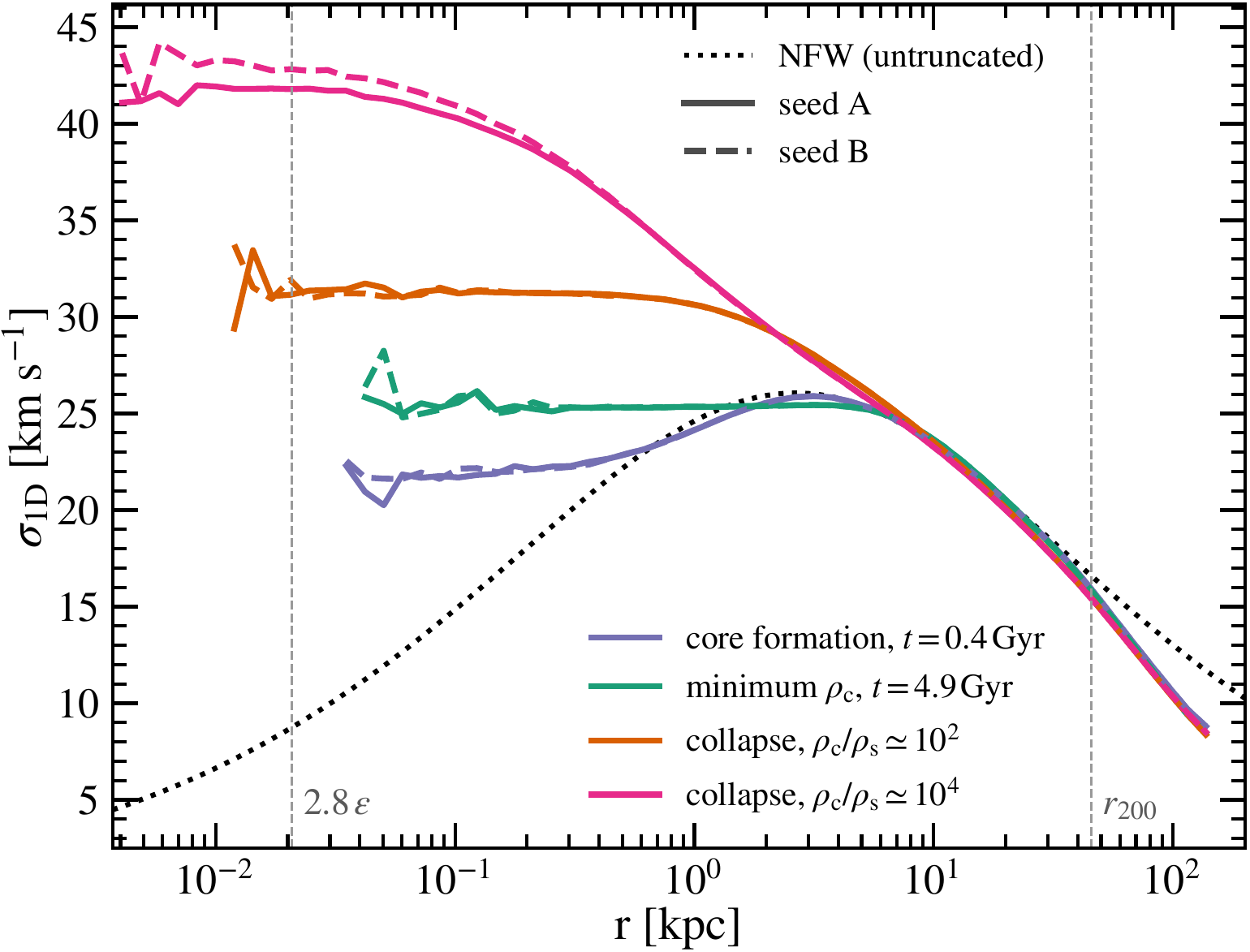}
    \caption{Reference solution with $N_{200}=10^8$ particles for two
    realizations of the scattering random numbers (seed A, solid; seed B,
    dashed). Top: central density as a function of time. The grey dotted
    line shows the fiducial $N_{200}=10^7$ simulation of
    \cref{subsec:coreCollapse} (O3, $\eta=0.005$), the circles mark the
    four epochs of the other panels, and the inset enlarges the final collapse.
    Middle and bottom: spherically averaged density and one-dimensional
    velocity dispersion profiles during core formation
    ($t=0.4\,\mathrm{Gyr}$), at the minimum of the central density
    ($t=4.9\,\mathrm{Gyr}$), and during the collapse at
    $\rho_{\rm c}/\rho_{\rm s}\simeq10^2$ ($t=29.5\,\mathrm{Gyr}$) and
    $\rho_{\rm c}/\rho_{\rm s}\simeq10^4$ ($t=30.5\,\mathrm{Gyr}$). For the
    last epoch, we show for each seed the snapshot closest to this density
    ($\rho_{\rm c}/\rho_{\rm s}=0.9\times10^4$ for seed A and
    $1.2\times10^4$ for seed B). The black dotted lines show the unsoftened, untruncated NFW density profile and its isotropic Jeans solution, and the vertical dashed lines mark $2.8\,\epsilon$
    and $r_{200}$.}
    \label{fig:reference}
\end{figure}

The top panel of \cref{fig:reference} shows the central density of both runs
as a function of time. The two realizations are indistinguishable until the
final collapse, and even there the times at which they reach a given central density differ by about $0.1$ per cent or less.
The fiducial $N_{200}=10^7$ run of \cref{subsec:coreCollapse} follows the same evolution but reaches a given central density about $0.1\,\mathrm{Gyr}$,
or $0.3$ per cent, earlier.

The lower panels of \cref{fig:reference} show the density and
velocity-dispersion profiles of both runs at the four epochs marked in the
top panel. We measure the profiles as in
\cref{subsec:collisionlessEquilibrium}. During core
formation, the self-interactions flatten the density profile inside about
$0.5\,\mathrm{kpc}$ and make the core isothermal. At $0.4\,\mathrm{Gyr}$, the core is still
colder than its surroundings, and the temperature inversion of the initial NFW
profile has disappeared only at the minimum of the central density. In the
collapse, the central density rises by almost four orders of magnitude above its
minimum, and the core becomes hotter than its surroundings. The profiles of
the two realizations agree at all four epochs. Only at the last epoch is a
difference visible.

\section{Limitations}
\label{sec:limitations}
\nori is designed for isolated self-interacting or collisionless dark matter systems, and the choices that make it fast narrow its range of application.

\subsection{Particle number}
Replicating the tree removes MPI communication from the tree walks, but requires the entire tree to fit into the memory of every compute node in CPU mode and every GPU in GPU mode. In the replicated GPU mode, the same holds for the particles, so that
adding GPUs does not raise the largest possible particle number. The
partitioned GPU mode divides the particles of one replica among the GPUs of a
compute node, which permitted the run with $1.47\times10^9$ particles in
\cref{subsec:scalingProblemSize}. However, a replica cannot extend over several
compute nodes, and every GPU still holds the complete tree. In CPU mode, the
limit is the shared memory of one compute node.

\subsection{Global time-steps}
The global time-step keeps all particles synchronized, removes asynchronous
scattering updates, and simplifies the load balancing. It becomes inefficient
when a small region sets the time-step for the whole system. During runaway
collapse, the global step shrinks as the gravitational or scattering time-step
limit of the core tightens, while the outer halo, which contains almost all
particles, is advanced far more often than it needs to be.
\Cref{subsec:performanceFullSimulation} quantifies the resulting reduction in the
speed-up over \arepoTwo.
Individual time-steps could be added without changing the replicated tree or
the division of the targets. With the current tree construction, a
straightforward implementation would still rebuild the whole tree at every
step, regardless of the number of active particles. This cost could be reduced
by reusing the tree over several steps and updating only the node moments
\citep{springel2001gadget,Miki2017GOTHIC}, or by a hierarchical gravity
scheme such as that of \textsc{gadget-4} \citep{springel2021simulating}.

\subsection{Physics}
Besides the restriction to isolated systems of equal-mass particles with
elastic, isotropic scattering (\cref{sec:nori}), \nori has no periodic or
cosmological boundary conditions and no external potential. Anisotropic or
dissipative interactions could be included by modifying the distribution of
$\hat{\bm{e}}$ and the magnitude of the outgoing relative velocity
\citep{TulinYuZurek2013SIDM,Robertson2017Anisotropic,Essig2019,vogelsberger2019evaporating,Shen2021dSIDM,Shen2022dSIDM,Rosenlyst2026}.

Some of these extensions would be straightforward. An external potential adds one
term to the acceleration. Periodic boundary conditions could be implemented
with Ewald summation \citep{Hernquist1991Ewald}, at the cost of additional computation
and storage during the tree walks.
Alternatively, a TreePM method could be used, which would also require a particle-mesh implementation on GPUs.
Unequal particle masses would require changes to the
scattering kinematics and the time-step criterion.

Anisotropic scattering can be included in the Monte-Carlo scheme. However, in these models, the differential cross-section peaks at small angles, and the momentum transfer cross-section can be much smaller than the total cross-section. Scattering events with small deflection angles change the velocities little, and resolving them forces a short time-step. Monte-Carlo schemes therefore can become unnecessarily expensive. For such models, \citet{Fischer2021} implemented a pairwise drag--diffusion scheme for the frequent small-angle scattering, and \citet{Arido2025HybridSIDM} combined it with Monte-Carlo sampling of the large-angle part. In the
drag--diffusion part, every particle interacts with \textit{all} of its neighbours in every time-step, and these interactions have to be synchronized. \nori gains much
of its efficiency on GPUs from the opposite situation, in which only a small
fraction of the trials is accepted, so that adopting the drag--diffusion scheme would
require a different organization of the scattering.

Inelastic scattering would change the outgoing relative speed of a pair.
\nori decides all trials of a step with the velocities before any event of
the step is applied, so later trials do not see the energy lost in earlier events of
the same step. For elastic scattering, we found no systematic effect of this error
in the comparison of \cref{subsec:coreCollapse}, which tests the approximation only indirectly. For inelastic scattering, every event cools the pair, and the
error becomes systematic, so the frozen-rate approximation would have to be
tested as a function of the time-step.

Finally, \nori does not yet include hydrodynamics or a galaxy-formation model. Both could in principle be added, but until then such simulations require a general-purpose code such as \arepoTwo \citep{ZierSidm2026}.

\section{Conclusions}
\label{sec:conclusions}

We have presented \nori, an $N$-body code for isolated systems of self-gravitating and self-interacting dark matter that runs on multi-core CPUs and GPUs. Its central design choice is to replicate the source data and distribute the walks over target particles. In CPU mode, each compute node shares one particle replica and one tree. In GPU mode, every GPU holds a full tree, while a particle replica occupies either one GPU or several GPUs within a node that access it through peer-to-peer reads. The gravity and SIDM neighbour walks therefore require no MPI communication. The memory required by the replicated data limits the code to around $2\times10^9$ particles.

\nori combines a deterministic tree built from Morton-sorted particles with one-sided monopole or quadrupole gravity and global kick--drift--kick time integration. Scattering probabilities are evaluated in parallel from a frozen state, and accepted events are applied in one common order, each using the velocities left by the events before it. This permits multiple events per particle while conserving each pair's momentum and kinetic energy to round-off. Fixed floating-point rules and reproducible random streams give bitwise-identical evolution across the CPU and CUDA configurations tested.

In \cref{sec:validation}, we tested \nori on an NFW halo with $10^7$ particles within $r_{200}$. Without self-interactions, the halo keeps its profiles over $50\,\mathrm{Gyr}$ outside the innermost $0.2\,\mathrm{kpc}$, where two-body relaxation lowers the density. The quadrupole runs have energy errors below $10^{-4}$ at the direct-summation epochs, the last at $49\,\mathrm{Gyr}$. With $\sigma/m=30\,\mathrm{cm^2\,g^{-1}}$, the gravothermal collapse agrees closely with our implementation in \arepoTwo, and the quadrupole runs show reduced energy errors in the final collapse compared with \arepoTwo with individual time-steps and with the monopole runs.

For the $N_{200}=10^8$ benchmark with global time-steps in both codes, \nori is 4 times faster per step than \arepoTwo on one 112-core CPU node and 7.7 times faster on 16 CPU nodes. One node with four A100 GPUs is 15 times faster than one CPU node running \nori and 60 times faster than one CPU node running \arepoTwo. The partitioned GPU mode allowed a benchmark with $1.47\times10^9$ particles on one GPU node. 

In the $N_{200}=10^7$ full-collapse comparison, where \arepoTwo uses individual time-steps, the GPU speed-up is 34 at $10^3\,\rho_{\rm s}$ and 25 at $10^4\,\rho_{\rm s}$. The advantage shrinks during the runaway collapse, when a small core sets the global time-step for the whole halo. This speed allowed us to follow the same halo with $N_{200}=10^8$ particles to a central density of $10^4\rho_{\rm s}$. For the two random-number realizations, the times at which a given central density is reached during the final collapse differ by about $0.1$ per cent or less.

Within its memory limit and its restriction to isolated, equal-mass systems with elastic, isotropic scattering (\cref{sec:limitations}), \nori makes high-resolution simulations of gravothermal collapse affordable on a single GPU node. We plan to use it for parameter studies of the collapse and for other isolated systems, such as dark-matter-only collisions of galaxy clusters, which require no changes to the code.

\section*{Acknowledgements}
We thank Volker Springel for useful discussions.
OZ acknowledges support from Harvard University through the Institute for Theory and Computation Fellowship.
The computations in this paper were run on the FASRC Cannon cluster supported by the FAS Division of Science Research Computing Group at Harvard University.
We used generative AI tools (e.g.\ ChatGPT, OpenAI, and Claude, Anthropic) for language editing and coding assistance. All AI-generated suggestions were critically reviewed, tested, and validated by the authors, who take full responsibility for the scientific content and results.

\section*{Data Availability}
The data underlying this article will be shared on reasonable request to the corresponding author (OZ).
We plan to make the \nori code publicly available in mid 2027. Until then, it is available on request.



\bibliographystyle{mnras}
\bibliography{main} 




\appendix
\crefalias{section}{appendix}
\section{Gravitational softening kernel}
\label{app:SofteningKernel}

We use the cubic-spline softening kernel of
\citet{springel2001gadget}. The potential generated by a
particle of mass $m_j$ is
\begin{equation}
    \Phi_{\epsilon,j}(r) = \frac{Gm_j}{h_\epsilon}\,W_2(u),
    \label{eq:softened_pair_potential}
\end{equation}
where $u=r/h_\epsilon$ and $h_\epsilon=2.8\epsilon$ is the radius beyond which
the force is Newtonian. We keep $h_\epsilon$ fixed for all particles;
it is independent of the adaptive SIDM smoothing length $h_i$.
The dimensionless softening kernel of the potential is
\begin{equation}
W_2(u)=
\begin{cases}
\displaystyle
\frac{16}{3}u^2-\frac{48}{5}u^4+\frac{32}{5}u^5-\frac{14}{5},
& 0\leq u<\frac{1}{2},\\[5pt]
\begin{aligned}[b]
&\frac{1}{15u}+\frac{32}{3}u^2-16u^3\\
&\quad+\frac{48}{5}u^4-\frac{32}{15}u^5-\frac{16}{5},
\end{aligned}
& \frac{1}{2}\leq u<1,\\[5pt]
\displaystyle -\frac{1}{u},
& u\geq1.
\end{cases}
\label{eq:softened_potential_kernel}
\end{equation}
At the origin, $W_2(0)=-14/5$, giving
$\Phi_{\epsilon,j}(0)=-Gm_j/\epsilon$. The parameter
$\epsilon$ therefore corresponds to the softening length of
a Plummer potential with the same central depth.

\begin{figure}
    \centering
    \includegraphics[width=\linewidth]{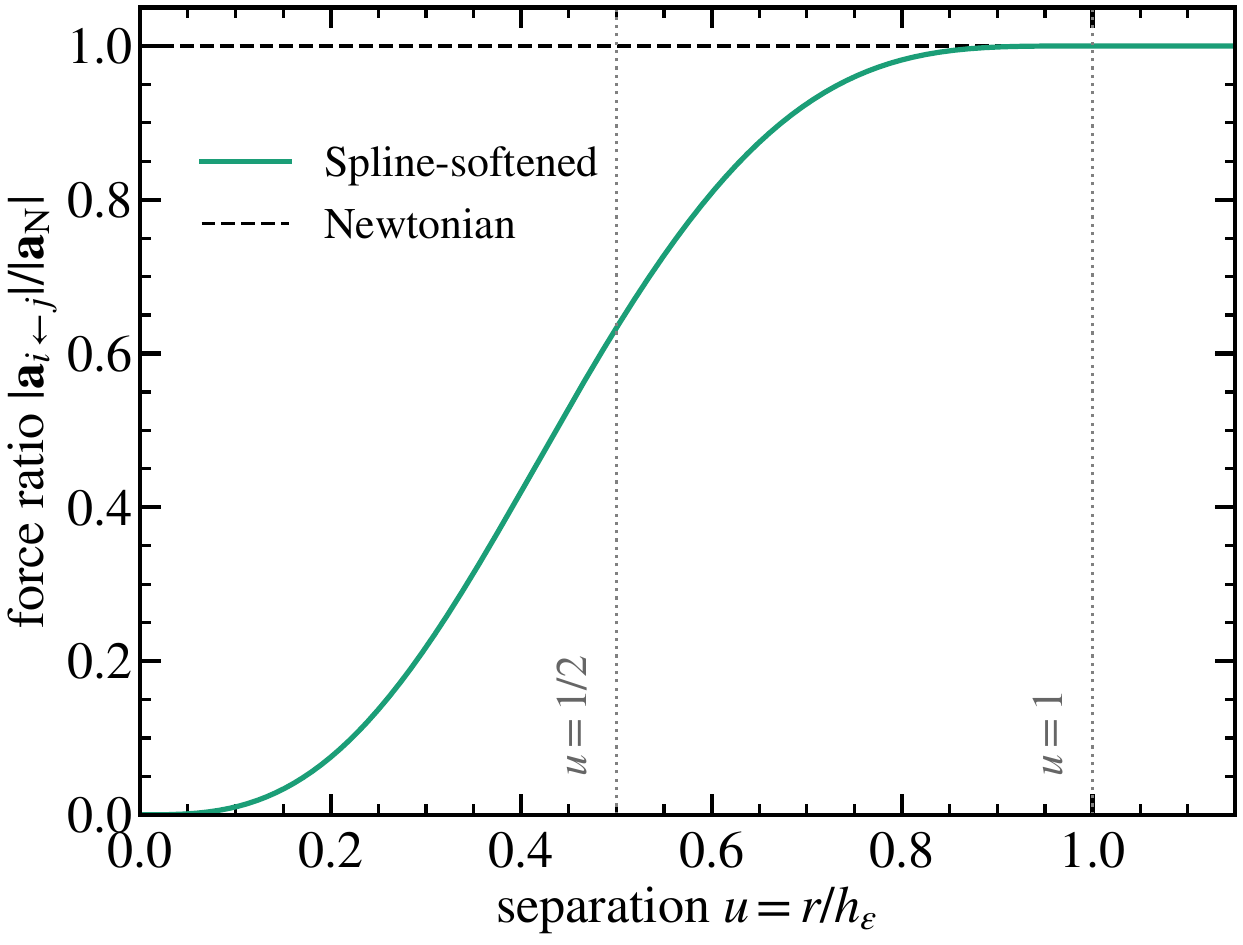}
    \caption{Force of the cubic-spline softening kernel relative to the Newtonian
    force as a function of $u=r/h_\epsilon$, with $h_\epsilon=2.8\epsilon$. The dotted lines mark the
    boundaries between the pieces of the kernel at $u=1/2$ and $u=1$.
    The ratio approaches zero as $u^3$ near the origin and
    is unity for $u\geq1$.}
    \label{fig:softening_force_comparison}
\end{figure}

We write the radial Green function of
\cref{eq:softened_potential} as
$g(r)=\varphi_\epsilon(r)=-W_2(u)/h_\epsilon$. Differentiation then gives the
softening kernel of the force,
\begin{equation}
    g_1(r)=\frac{g'(r)}{r}
    =-\frac{W_2'(u)}{h_\epsilon^3u}.
    \label{eq:softened_g1_definition}
\end{equation}
Its explicit form is
\begin{equation}
g_1(r)=\frac{1}{h_\epsilon^3}
\begin{cases}
\displaystyle
-\frac{32}{3}+\frac{192}{5}u^2-32u^3,
& 0\leq u<\frac{1}{2},\\[5pt]
\begin{aligned}[b]
&\frac{1}{15u^3}-\frac{64}{3}+48u\\
&\quad-\frac{192}{5}u^2+\frac{32}{3}u^3,
\end{aligned}
& \frac{1}{2}\leq u<1,\\[5pt]
\displaystyle -\frac{1}{u^3},
& u\geq1.
\end{cases}
\label{eq:softened_force_kernel}
\end{equation}
For the pair separation $\bm{r}_{ij}=\bm{x}_i-\bm{x}_j$,
with $r_{ij}=|\bm{r}_{ij}|$, the acceleration of particle
$i$ due to particle $j$ is then
\begin{equation}
    \bm{a}_{i\leftarrow j}
    =Gm_jg_1(r_{ij})\,\bm{r}_{ij}.
    \label{eq:softened_pair_acceleration}
\end{equation}
The finite limit $g_1(0)=-32/(3h_\epsilon^3)$ gives zero acceleration
at zero separation and a harmonic force for $r_{ij}\ll h_\epsilon$. The quadrupole interactions of O3
described in \cref{subsec:gravityImplementation} also require the second and third
radial derivatives of $g$, which follow from
\cref{eq:softened_force_kernel} by differentiation and reduce to those of the
Newtonian potential for $r\geq h_\epsilon$. For non-zero separations, the
ratio of the softened to the Newtonian force is
\begin{equation}
    \frac{|\bm{a}_{i\leftarrow j}|}{|\bm{a}_{\rm N}|}
    =-r^3g_1(r),
    \label{eq:softened_force_ratio}
\end{equation}
which is shown in \cref{fig:softening_force_comparison}.

\section{Impact of tree-based potential-energy estimates on energy diagnostics}
\label{app:ImpactEnergyMeasurement}
The energy plots of \cref{sec:validation} use potential energies from direct summation in post-processing. \nori also records the energy during the run, but, as in \arepoTwo and \gizmo, its on-the-fly potential energy uses the same approximate tree method as the force calculation. This approximation can increase the scatter in the measured energy evolution. Moreover, if the tree origin is not randomized, temporally correlated errors in the potential energy can produce an artificial secular evolution.

\begin{figure}
    \centering
    \includegraphics[width=1\linewidth]{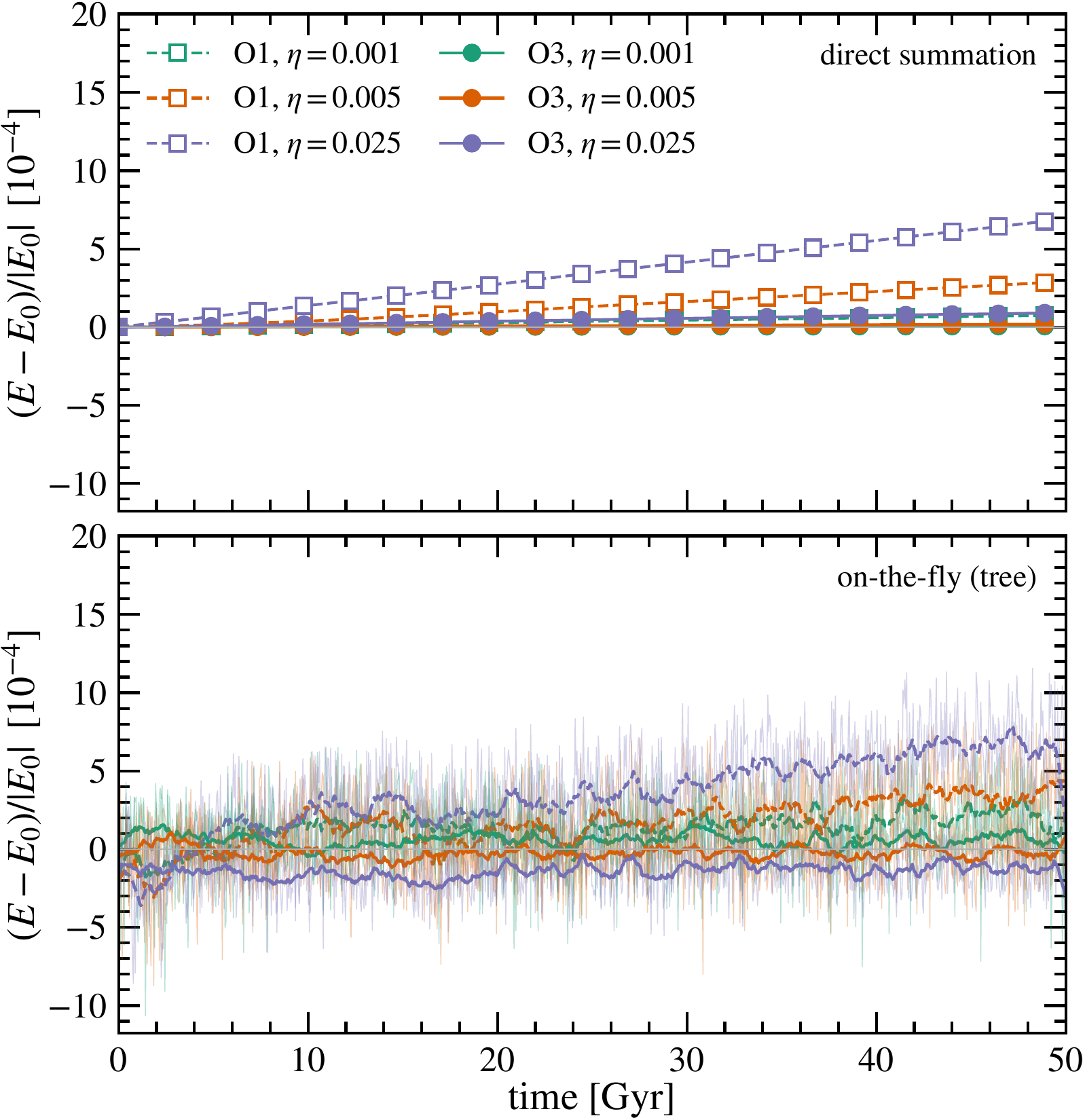}
    \caption{Relative change in the total energy of the collisionless
    simulations of \cref{subsec:collisionlessEquilibrium} for two
    estimates of the potential energy, in units of $10^{-4}$. Colours and
    line styles are those of \cref{fig:collisionless_1e7_energy}. Top:
    direct summation over all particle pairs for the initial conditions and
    20 snapshots, as in \cref{fig:collisionless_1e7_energy}; $E_0$ is the
    energy of the initial conditions from the direct sum.
    Bottom: the on-the-fly statistics, which use the tree; here $E_0$ is the median
    of the tree energy over the first $0.5\,\mathrm{Gyr}$, which suppresses the
    start-up noise. Thin lines show
    every record (one per $0.05\,\mathrm{Gyr}$), thick lines their running
    mean over $1\,\mathrm{Gyr}$.
    }
    \label{fig:energy_direct}
\end{figure}

\Cref{fig:energy_direct} compares both estimates for the collisionless
simulations of \cref{subsec:collisionlessEquilibrium}. With direct
summation, the energy error grows smoothly with time and increases with
$\eta$ for both multipole orders. The on-the-fly estimate scatters around its
running mean with a standard deviation of $1.5\times10^{-4}$ for O3 and
$2.5\times10^{-4}$ for O1, which exceeds the secular change of all O3 runs,
and the running mean itself still fluctuates by more than $10^{-4}$. For O3
with $\eta=0.025$, it indicates an energy loss of
$(1$--$3)\times10^{-4}$, whereas the energy increases by $9\times10^{-5}$.
We therefore argue that studies of energy conservation during gravothermal collapse should generally validate tree-based energy estimates against direct-summation calculations.


\bsp	
\label{lastpage}
\end{document}